\documentclass[pr,amsmath,amssymb,superscriptaddress]{revtex4}
\usepackage{graphicx,xcolor}
\usepackage{hyperref}
\usepackage{bm}

\usepackage{epsf}
\usepackage{amsmath,amsfonts}

\newcommand{\la}{\label}
\newcommand{\bea}{\begin{eqnarray}}
\newcommand{\eea}{\end{eqnarray}}
\newcommand{\beq}{\begin{equation}}
\newcommand{\eeq}{\end{equation}}
\newcommand{\be}{\begin{equation}}
\newcommand{\ee}{\end{equation}}
\newcommand{\ii}{{\rm{i}}}
\newcommand{\dd}{{\rm{d}}}

\newcommand{\p}{\partial}
\newcommand{\nn}{\nonumber}

\usepackage{color}

\begin{document}

\title{Shocks and  finite-time singularities in Hele-Shaw flow}

\author{S -Y Lee}
\address{Centre de recherches math\'ematiques, University of Montreal, Montreal, Canada}
\author{R Teodorescu} 
\address{Center for Nonlinear Studies and T-4, Los Alamos National Laboratory, Los Alamos, NM 87505, USA}
\author{P Wiegmann}
\address{The James Franck and Enrico Fermi Institutes, University of Chicago, 5640 S. Ellis Ave, Chicago IL 60637, USA}

\begin{abstract}
Hele-Shaw flow at vanishing surface tension is ill-defined. In finite time, the  flow develops cusp-like singularities. We show that the ill-defined problem admits a weak {\it dispersive} solution when  singularities  give rise  to a graph of shock waves propagating in the viscous fluid. The graph of shocks grows and branches.   Velocity and pressure jump across the shock. We formulate a few simple physical principles which single out the  dispersive solution and interpret shocks as lines of decompressed fluid. We also formulate the dispersive weak solution in algebro-geometrical terms as an evolution of the Krichever-Boutroux complex curve. We study in detail the most generic (2,3) cusp singularity, which gives rise to an elementary branching event. This solution is self-similar and expressed in terms of elliptic functions.

\end{abstract}


\maketitle

\section{Introduction}

The zero surface tension limit of Hele-Shaw flows \cite{Hele-Shaw}  describes a planar interface between two incompressible and immiscible phases propagating with velocity equal to the density of the harmonic measure of the interface.  It is a famous problem whose importance goes far  beyond applications to fluid dynamics.  Originally formulated by Henry Darcy in 1856 \cite{Darcy} in relation with groundflow of water through porous soil,  it describes a large  class of two-dimensional growth processes driven by a harmonic field (a.k.a Laplacian growth) and  has led to a number of important developments in mathematics and mathematical  physics,  from complex analysis to random matrix theory.

\begin{figure}[ht!]
\begin{center}
\includegraphics*[width=8cm]{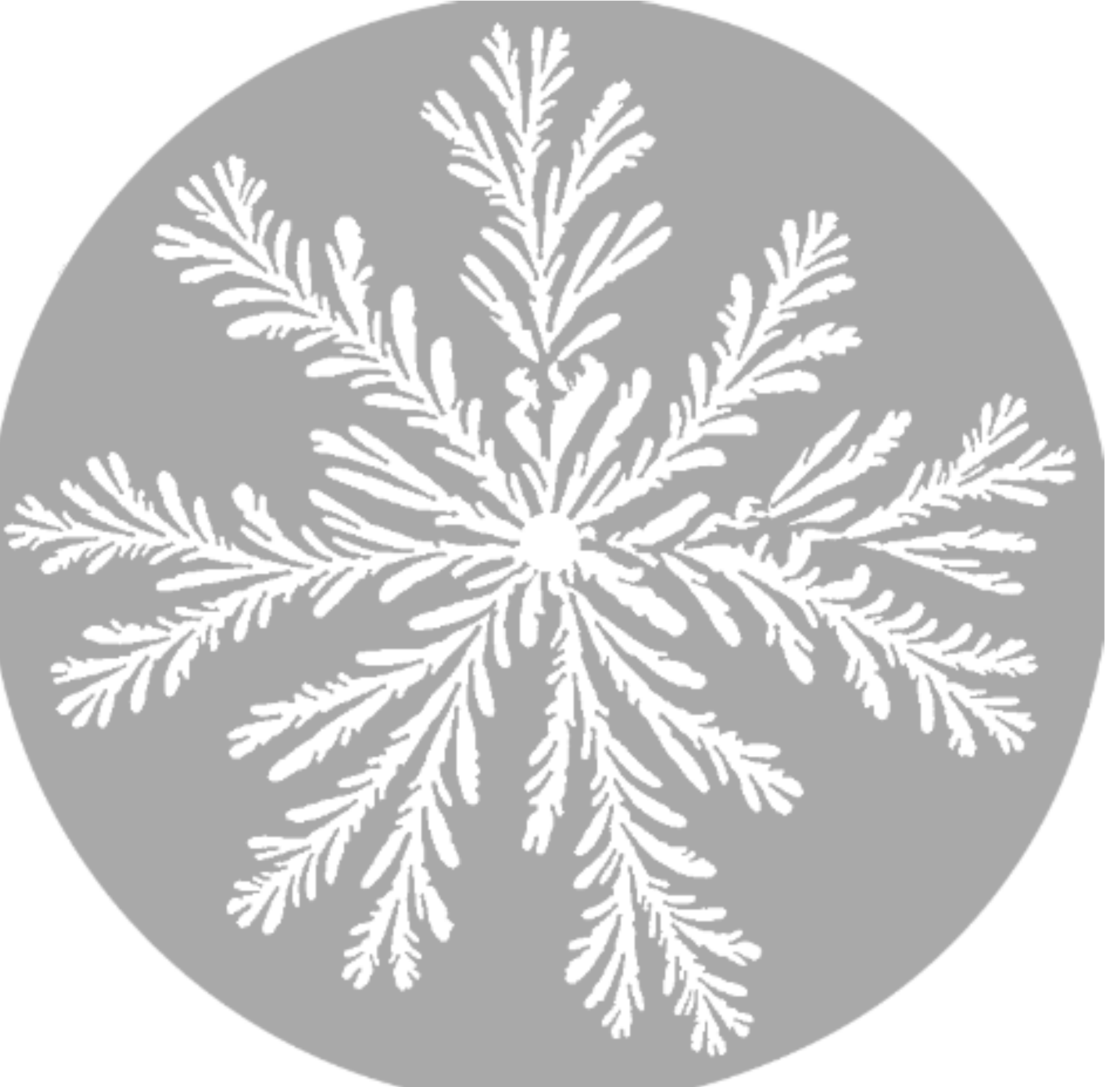}
\includegraphics*[width=8cm]{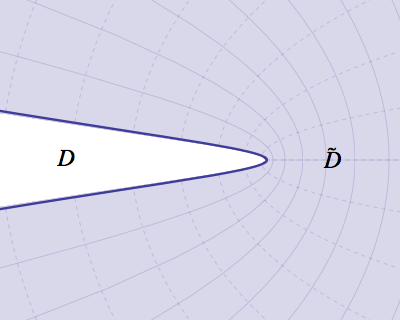}
\caption{(Left) Experimental pattern in Hele-Shaw cell \cite{Sw}. The shaded region in both pictures  is the viscous fluid. (Right) Viscous fingering in Hele-Shaw flow. Solid  lines are lines of  a constant pressure, dashed lines are stream lines.}
\label{experiment}
\label{suction}
\end{center}
\end{figure}

A compact formulation of the problem is:  let $\gamma(t)$ be a simple planar curve  -- boundary of a simply-connected domain $D$. The curve evolves in time $t$ according to 
\be\la{D}
{\rm{Darcy\; Law:} }\quad v_n(z)\sim H(z),\quad z\in \gamma(t).
\ee
Here $v_n$ is the velocity of the curve - an outward normal vector from $D$, and  $H(z)$  is the normal gradient of a solution of the Dirichlet problem in a fluid domain $\widetilde D$ with a source (a sink) at a distant location (often at infinity in theoretical setup; not shown in Figure~\ref{suction}):
\begin{align}\la{D0}
&H(z)=-\p_n p(z), \\
\la{D1}
&\Delta p=0\;\;\mbox{ on } \tilde D,\quad p_{|_\gamma}=0,\quad p_{ |_{z\to\infty}}\sim - \log|z|.
\end{align}
We use $H(z)|\dd z|$ for the harmonic measure on the curve $\gamma(t)$. 

\begin{figure}[htbp]
\begin{center}
\includegraphics[width=8cm]{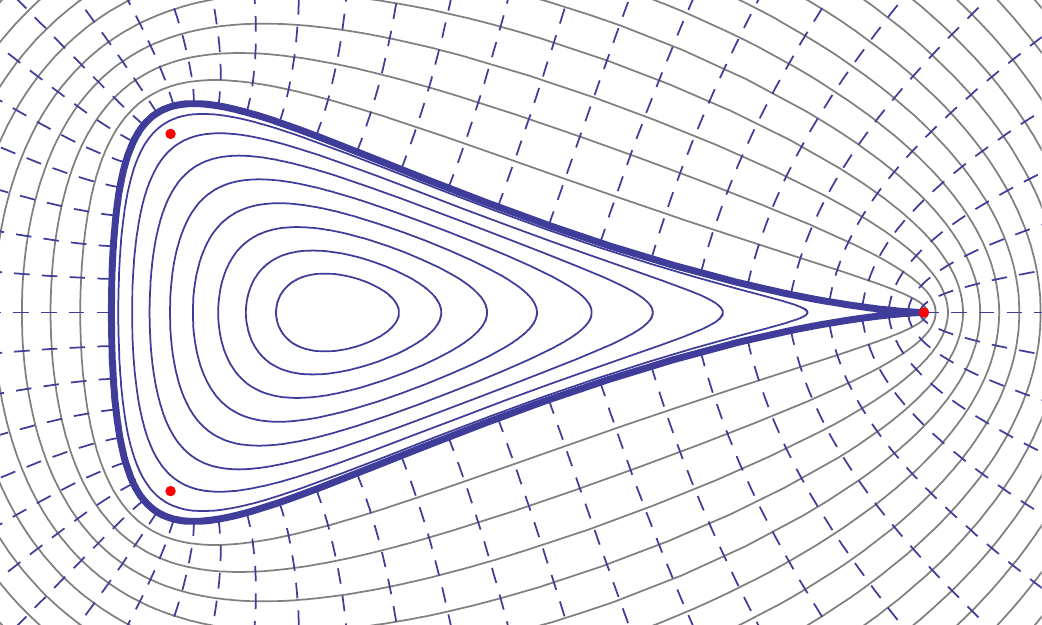}$\quad$
\caption{Deformed hypertrochoid 
$z(w)=r\,w+\frac{A_r}{2r}+\frac{A_r}{w}+\frac{r^2}{w^2}$ with $A_r=r-4r^3-\sqrt{\frac{3r^2}{5}-8r^4+16 r^6}$, 
 where one critical point gives rise to a (2,3) singularity while two other critical points remain in the domain $D$.  Closed lines inside the critical boundary show the pre-critical evolution.}
\label{dhypo}
\end{center}
\end{figure}

Laplacian growth models \cite{Hohlov-Howison94, Gustafsson-Vasiliev06} are
characterized by intricate finger-like unstable  patterns \cite{ST}  featuring  finite-time singularities  \cite{bs84, Howison85} in the form of boundary cusp formation (see e.g. Figures~\ref{shypo}, \ref{dhypo}).

As such the problem is ill-defined: the Darcy law stops making sense, since velocity and pressure gradient diverge at a {\emph {critical time}} in a {\emph {critical point}}, where a cusp-like  singularity occurs.

An important feature of Laplacian  growth is its integrability. In this case, integrability means that, if the initial interface is  an algebraic curve of a given order,  it will remain so at all times before the critical time. Insights into integrability appeared in early papers \cite{Richardson72} and then were further developed in more recent studies \cite{us1, us3, us4, us5, us6}.

Dynamics near a critical point (i.e., close to the blow-up time) belongs to the class of non-linear problems for which any perturbation is singular.  In this case, various regularization schemes typically lead to different results.

This situation is typical in fluid mechanics and, in fact,  similar to singularities appearing in  compressible  Euler flows. There, perturbing the Euler equations by a diffusive mechanism  (i.e., by viscosity), or by a Hamiltonian mechanism (i.e., by dispersion) leads to  different physics and different flow patterns  (as an example compare the dissipative Burgers equation and the Hamiltonian KdV equation, which differ by terms with higher-order derivatives).

A traditional  laboratory set-up (called a Hele-Shaw cell after its inventor \cite{Hele-Shaw}) consists of two horizontal  plates separated by a narrow gap, initially filled with a viscous liquid, where inviscid liquid  is pumped in at a constant rate  at the center of the cell, pushing the viscous liquid, away to a sink. Below we will refer to the viscous liquid  as {\it fluid}, occupying a domain $\tilde D$. An  inviscid liquid occupies a complimentary domain   $ D=\mathbb{C} \setminus \tilde D$. 
Both liquids are considered to be incompressible. The  boundary initially moves according to the Darcy law (\ref{D}-\ref{D1}), where $p$ is the pressure of the
incompressible viscous fluid.  A typical pattern is seen in Figure~\ref{experiment}.

It seems that the  most relevant mechanism of taming singularities  for experiments in Hele-Shaw cell (even for  a developed pattern as in Figure~\ref{experiment}
\cite{Sw}) is surface tension.  Then $p$ in (\ref{D1}) along the interface is proportional to its curvature. This perturbation stabilizes growing patterns (as seen in Figure~\ref{suction})  but destabilizes their analytical and integrable  structure. Also, it is not expected that patterns possess universal features if the surface tension is significant.

The  surface tension effects  can be diminished in non-Newtonian fluids where a viscous fluid is substituted by a polymer solution. In this case, singularities are controlled by certain visco-ellastic  properties of the fluid \cite{Maher}.

An interesting modern  setting consists only of the viscous liquid without the inviscid liquid. In experiments described in \cite{Lipson}, a thin layer of viscous liquid  is positioned on a horizontal wetted substrate,  with the free boundary contracting by some suction mechanism. \footnote{In experiments of Ref.~\cite{Lipson},  the liquid evaporates uniformly including the boundary. Here we  assume that  suction is applied to a point of the liquid away from its boundary.}The advantage of this setting is that the height of the layer may vary (and indeed does vary along the rim \cite{Lipson}), which effectively makes the two-dimensional flow compressible. Below we assume such setting that allows the variation of compressibility, and will keep the name {\it Hele-Shaw flow}.

Another interesting setting occurs  when air propagates through granular media \cite{Jag-Nag}. There, the boundary is not controlled by its curvature (unlike in fluids with surface tension), but the density of granular media may vary on the boundary.

The traditional approach in studies of  Hele-Shaw flow assumes the following order of limits:  the density of  the fluid  is set to a constant  first,  while the surface tension is kept non-zero. Then  the limit of surface tension  going to zero is taken, but proves to be  ill-defined; the solution blows up.

This requires a different model for Laplacian growth,  such that the  zero surface tension limit \eqref{D0} is well-defined and does not lead to a singular solution.  Such a  view for Laplacian growth is suggested by the setting with a thin layer on a wetting substrate \cite{Lipson}, or propagating air through  granular media \cite{Jag-Nag} mentioned above, and also by  the model of Diffusion-Limited Aggregation (DLA) \cite{DLA81}, or iterative conformal maps algorithm \cite{HastingsLevitov}. In the latter case, the interface evolves by coalescence (aggregation) of Brownian walkers supplied at a constant rate from a point source (infinity). The probability of aggregation at a given point is proportional to the harmonic measure of the existent aggregate.  In the limit when the area of a  particle  tends to zero, the problem of aggregation becomes identical to  the ill-defined zero surface tension Hele-Shaw flow. Aggregation continues beyond this limit, producing notoriously intriguing fractal patterns. In this case, the zero-size limit for particles is  a singular perturbation.   Due to a finite size and irregular shape of particles, taking the limit of aggregation model to the continuous fluid mechanics effectively leads to a compressible fluid, as in the two settings  mentioned  previously.

The  geometric regularization suggested by aggregation models  is a primary motivation of this paper (and of preceding papers  \cite{us1, us3, us4, us5, us6}). 
We will relax  conditions that  fluids are incompressible and curl-free at a microscale, but will require that liquids are curl-free at a larger scale and then study a singular  limit when compressibility vanishes, setting surface tension to zero in the first  place. 
  
Based on this idea, we will construct a {\it weak solution} of the Hele-Shaw zero-surface tension flow, where:
\begin{itemize}
\item[-]
formation of a  cusp singularity is followed by expanding {\it shocks} -- one dimensional lines through which  pressure and velocity of the fluid are discontinuous.   The density of the fluid (which is constant away from the shock) becomes singular at the shock, which  can be interpreted as a deficit of  fluid;
\item[-] shock lines form an evolving branching pattern. The geometry of shocks is governed by transcendental equations. 
 \end{itemize}
\begin{figure}[htbp]
\begin{center}
\includegraphics[width=7cm]{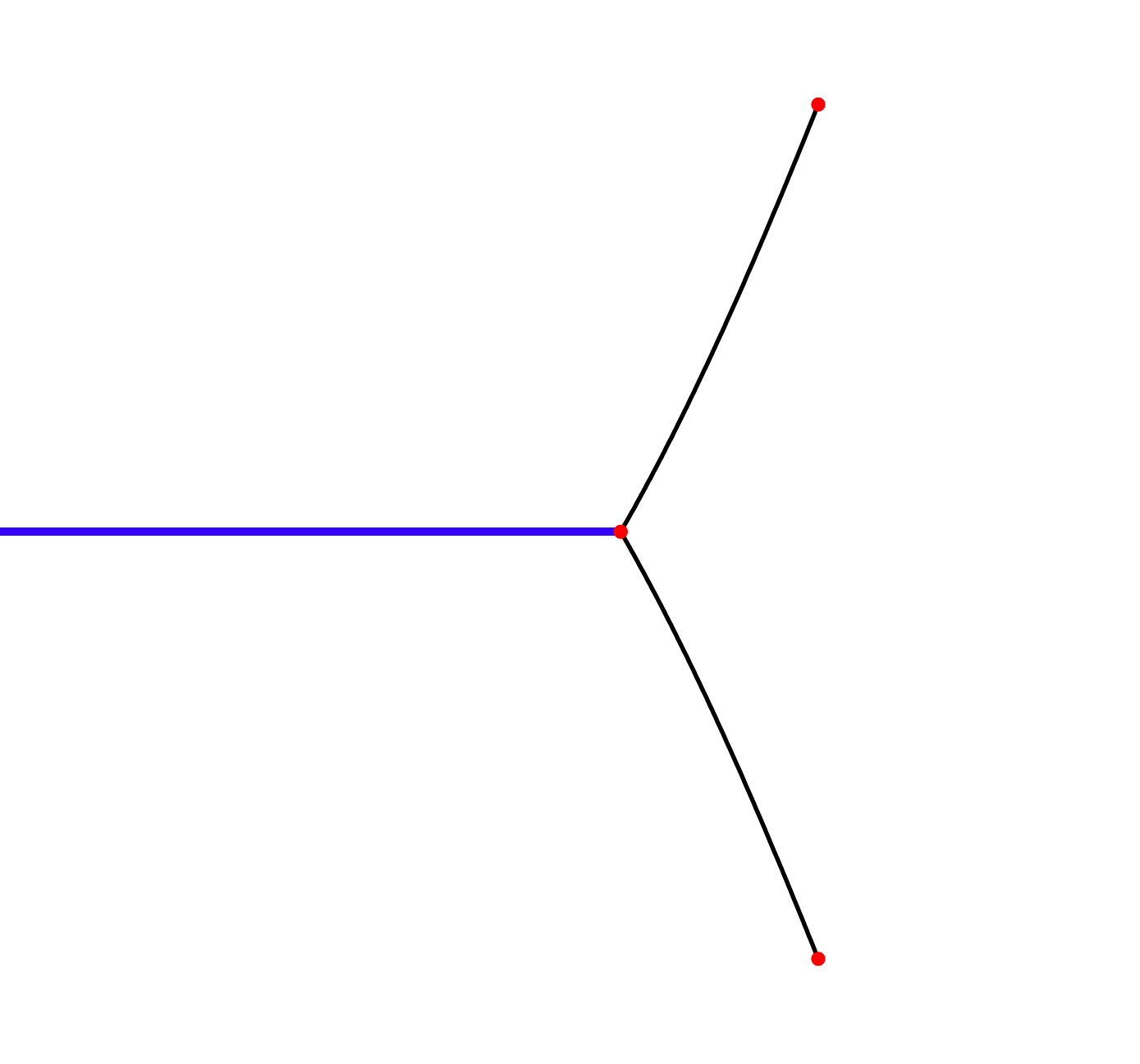}$\quad$
\includegraphics[width=7cm]{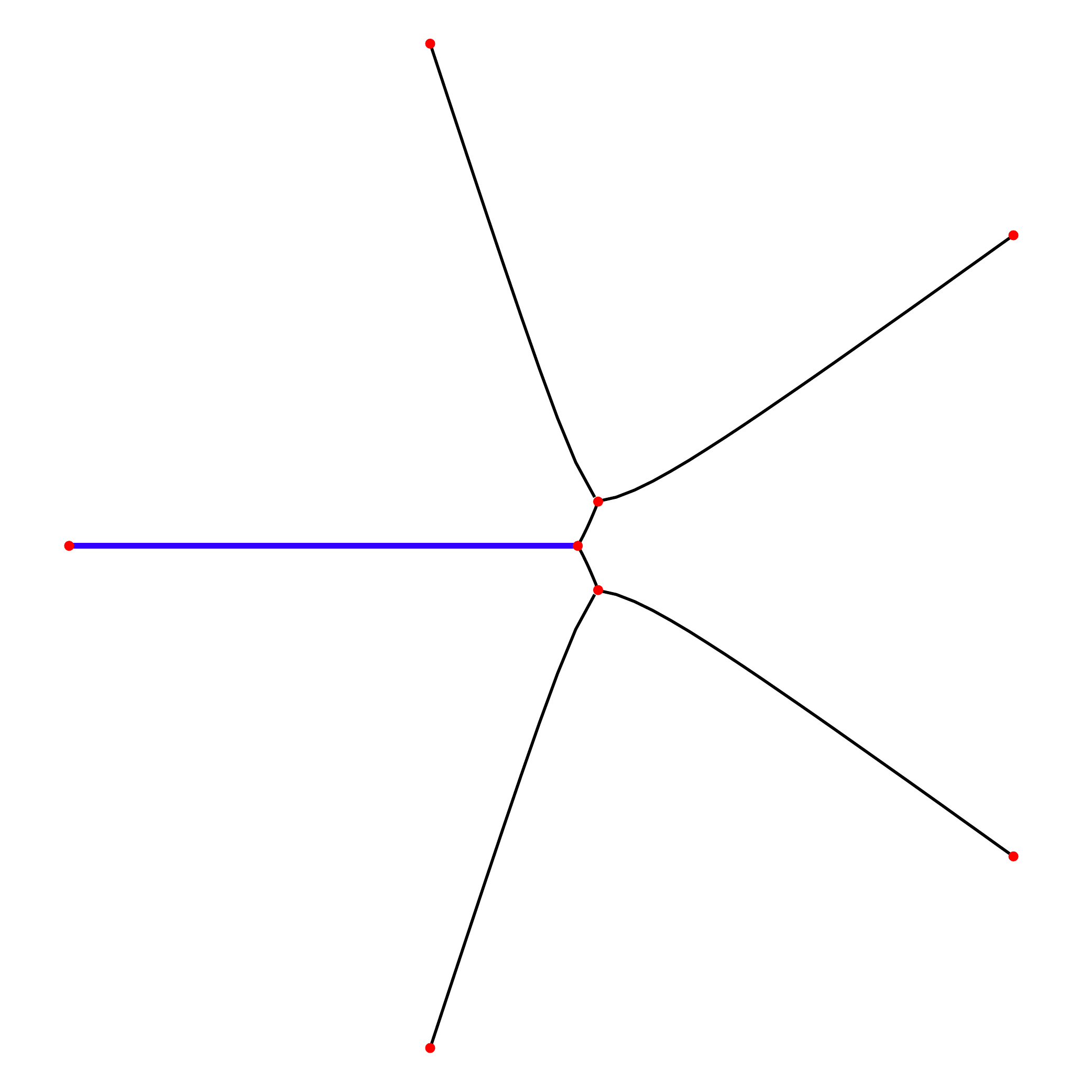}
\caption{A growing and branching shock pattern, with one (left) and two (right) generations of branchings. The bold  line along the negative $x$-axis represents a narrow viscous finger (fluid). At this scale, the viscous finger is vanishingly narrow. 
}
\label{shocks}
\end{center}
\end{figure}

In a subsequent publication,  
we will  describe  connections between weak solutions of the Hele-Shaw problem and a few related mathematical problems.  We will show that:
\begin{itemize}
\item[-]
the lines of discontinuities (shocks) correspond to accumulation of zeros of  bi-orthogonal polynomials of large orders;
\item[-]
 the shock fronts are anti-Stokes lines for the isomonodromy problem naturally related to the Whitham averaging of solutions of Painlev\'e equations corresponding to the integrable models associated with particular types of cusps.
  \end{itemize}

  Shock fronts typically form a growing, branching tree, as in Figure~\ref{shocks}. In this paper, we provide a detailed analysis of the shock pattern  solution representing the birth of the  first branching event  (Figure~\ref{shocks},  above) of what
  will become a complicated degree-two tree. This solution is self-similar and  represents the local  branching structure  of a developed tree. 

A comment is in  order: typically, shock waves in hydrodynamics are associated with phenomena occurring at high Reynolds numbers, due to the inertial term in Euler equations \cite{LL}.  Here we encounter a new situation when a discontinuous solution occurs in a viscous flow where the Reynolds number is small and inertia terms are neglected in the Navier-Stokes equations (see the next section). For a discontinuity of this kind, perhaps  ``crack" rather than ``shock"  may be a better name.

The paper  consists of three parts: 
first we present a few (necessary) known facts about singularities in the Hele-Shaw flow and formulate the problem in terms of {\it inverse balayage} -- an important concept giving insights to the weak solution. Then we formulate the  {\it weak dispersive solution} which allows  shocks, and  describe their hydrodynamics (the  Rankine-Hugoniot condition). Next we formulate the flow in terms of evolution of a complex curve.  This formulation is the most suitable for computations.  Finally, we give a detailed analysis of a generic (2,3) -- cusp singularity and  describe the elementary branching event. This particular solution  reflects the nature of a more general weak solution. 

We start by formulating the differential and weak forms of zero surface tension  Hele-Shaw flow.

\section{Differential and weak forms of Hele-Shaw flow}
\subsection{Differential form of Hele-Shaw flow}\la{N}
A thin layer of viscous fluid of height $b$  occupies an open domain $\tilde D$ with a smooth simply-connected  boundary $\gamma(t)$. The fluid is drained from a point  $z=\infty$ at a constant rate $Q$,  Figure~\ref{experiment}. 

We assume a quasi-stationary flow  and therefore neglect  the term $\frac{\partial \mathbf{v}}{\partial t} $  in  the Navier-Stokes equation
$\rho\left(\frac{\partial \mathbf{v}}{\partial t} + \mathbf{v} \cdot \nabla \mathbf{v}-\nu \nabla^2 \mathbf{v}\right) = -\nabla p ,$
 where $\nu$ is the kinematic viscosity, and $\rho$ the density of the liquid. We also  assume that the Reynolds number is small $Re =  |\mathbf v| b/\rho\nu \to 0$, and further neglect the inertial term there. We obtain 
$\rho \nu \nabla^2 \mathbf{v} = \mathbf{\nabla} p$.
Further assuming a Poisseulle profile for the flow,  we replace  $\nabla_z^2\mathbf{v}$ by its average  $ -(12/b^2) \mathbf{v}$  and neglect $(\nabla_x^2 +\nabla_y^2 )\mathbf{v}$, to obtain the Darcy law
\be \la{darcy-cell}
\rho\mathbf{v} = -K \mathbf{\nabla} p,
\ee
where $K=\frac{b^2}{12 \nu}$  is called {\emph{hydraulic conductivity}}.  Evidently, this approximation is consistent  if the fluid is irrotational $\mathbf{\nabla}\times   \mathbf{j}=0$ , where $ \mathbf{j}=\rho \mathbf{v}$, as it is seen from the last formula. If furthermore the density is assumed to be spatially uniform, incompressibility of the fluid  $\mathbf{\nabla\cdot v} = 0$ and (\ref{darcy-cell}) imply that  pressure is  harmonic at constant density:
\be \la{oil}
\Delta p = 0.
\ee
 We will relax the condition of constant density when defining weak solutions. The fluid is sucked out at a constant rate $Q=\oint_\gamma\mathbf{j} \times \dd {\mathbf l}$, where $\gamma$ is oriented counter-clockwise. According to the Darcy law (\ref{darcy-cell}), $Q$  equals $-K\oint _\gamma \mathbf{\nabla} p \times \dd \mathbf {l}$. Therefore, at the drain: 
 \be\la{s}p\to - \frac{Q}{2\pi K }\log\frac{|z|}{r(t)}+{\cal O}\left(\frac 1z\right) \; \mbox{ as } z \to \infty\ .
 \ee

If the boundary is smooth, the pressure on the boundary is controlled by the  surface tension $p=\sigma\times{\rm curvature}$.  If surface tension vanishes, the pressure is constant along the boundary $p=0$. Then the pressure is  a solution of the Dirichlet problem in the fluid.

The time dependent function $r(t)$ (known as conformal radius), or capacity ${\cal C}(t)=r(t)
$ \cite{Ahlfors}, where  $R$ is a cell radius, are   important characteristics of the flow. They monotonically grow and have  the following hydrodynamic interpretation \cite{Artem}: let us compute the power required to drain the fluid: 
$N(t) = -\int_{\tilde D} \mathbf{\nabla} p\cdot \mathbf{j}\, \dd x \dd y .
$
Darcy law  yields that the  power is the Dirichlet integral
$
N(t)  =  K \int_{\tilde D} (\mathbf{\nabla} p)^2 \dd x \dd y
$
evaluated as $N(t) =Q^2/(2\pi K)(\log R-\log {\cal C}(t)).$
Then the power is a monotonically decreasing function of time.    

We will see that, as a flow approaches a (2,3)-cusp singularity and enters into a shock regime, capacity increases non-analytically, as a square root of time measured from the critical point.  After a singularity,   capacity again increases 
as a square root of time measured from the critical point.  A notable fact is that  the coefficient of the square root singularity changes  discontinuously  before and after the transition  (see Sec.~\ref{capacity1}). 
Physically, it means that the power  required to produce the flow goes through a discontinuous change every time the shock front branches, such that more power is required than it would be in the absence
of singularity formation  \cite{Artem}.

\subsection{Weak form of Hele-Shaw flow} \la{section_three}
Once  the flow reaches a  cusp singularity, it cannot be continued any longer. The Darcy law written in differential form loses it meaning  at the cusp -- harmonic measure there diverges. 

This situation is typical for differential equations \cite{Evans}  with conservation laws of {\emph{hyperbolic type}}:
\be \la{cauchy}
\p_t {{u}} + \partial_z f(u) = 0.
\ee

Cauchy problems  with such conservation laws are ill-posed, and  smooth initial data evolve into  singularities developing in finite time (such as shocks, vortices, etc.).  Darcy equation  is of this type.  

The origin of this phenomenon is that conservation equations of hyperbolic type are approximations of well-defined problems. A deformation of these equations by  terms with higher gradients, controlled by a small parameter $\hbar$,  prevents formation of singularities. If a  smooth solution of a deformed equation ${ u}_\hbar$ leads to a space-time discontinuous function 
${u}|_{\hbar\to 0}$ as the deformation is  removed, it is  called a weak solution.  A discontinuity (a shock front) travels with a velocity given by the Rankine-Hugoniot condition - a weak form of the differential equation (\ref{cauchy})
 \cite{LL, Lamb}:
\be\la{RH1}
V=\frac{{\rm disc}\,f}{{\rm disc}\, u}.
\ee

In most cases, simple physical principles determine a weak solution  without actually  specifying a regularizing deformation. The best known example is the Maxwell rule determining the  position of shock front \cite{LL}, or more generally,  the Lax-Oleynik  entropy condition for a {\it viscous solution} of  equations with hyperbolic  conservation laws \cite{Evans}.  

Here we assume the same strategy. We will be looking for a weak  solution of the Hele-Shaw  problem when the Darcy law is applied everywhere in the fluid except a moving, growing and branching graph
$\Gamma(t)\subset \tilde D$  of shocks  (or cracks), and where  pressure  suffers a finite discontinuity -- a weak form of singularity. 

Below we formulate a few simple, natural physical principles which will guide us to obtaining a unique weak  solution, that  we call {\it a dispersive solution}.

Shocks may have  different  physical meanings depending on the various experimental settings discussed in the Introduction. In the two-liquids setting (traditional Hele-Shaw cell), shocks are narrow, extended channels where the inviscid fluid is  compressed and carries vorticity. In the single fluid setting, shocks are areas with a  deficit of  fluid.  Shocks communicate with the  bulk of the fluid by supplying/removing fluid to/from  the bulk.    Obviously, for shocks to occur, the uniform fluid density condition must be relaxed.  A difference between shocks and a regular boundary is  that both  pressure and velocity have steep gradients across a shock. Also, the pressure is generally  not a constant  along a shock line.

In the following,  we summarize the weak form of Hele-Shaw problem, postponing the details to 
Sec.~\ref{section_three}. The  weak form reads: 
\smallskip

\noindent - the differential Darcy law (\ref{darcy-cell}-\ref{s}) holds everywhere, except on the graph of shocks,  where pressure jumps and the density  (a constant everywhere else) has a single layer density deficit $\rho(\mathbf{x},t)=\rho_0-\delta({\bf x};\Gamma) \sigma({\bf x},t)$  with a  line density: $\sigma({\bf x},t)>0$ (here $\delta({\bf x};\Gamma)$ is the delta-function on a shock). Both pressure and line density vanish at  shock's endpoints ${\bf e}_k$ as $\sqrt{|{\bf x-\bf e}_k|}$. Shocks move with velocity whose normal component ${\bf V_\perp}$ is directed  towards  higher pressure regions,  and obeys the Rankine-Hugoniot condition:
\be\la{RH0}
\sigma |{\bf V}_\perp|=K|{\rm disc}\; p|. 
\ee
We discuss the motivation, interpretation and consequences  of this solution in  sections below.  The conditions imposed are  restrictive. They determine a  shock pattern with a given number of legs, up to a finite number of deformation parameters.  Shock graphs with one and two branching  generations are represented in Figure~\ref{shocks}.

\section{Finite-time singularities in Hele-Shaw flow}

The dynamics described by the Darcy law (\ref{D}, \ref{D0})  is
ill-defined. This has been understood  from different angles
\cite{ST, Richardson72, PK}, and can be seen from a simple argument:
let $w(z)$ be a conformal univalent map of  the fluid domain
$\tilde D$ onto the exterior of the unit circle, $|w| > 1$; the solution of the
exterior Dirichlet problem then reads
\be p(z)= -
\frac{1}{2} \log |w(z)|, \quad 
v=\frac{1}{2} |w'(z)|,\;z\in {\tilde D}. 
\ee 
(From now on we set $K=1,Q=\pi$. In these units the area is  equal to $\pi\times \rm{time}$. In this section we also set $\rho=1$).
In other words,
velocity  of a boundary point is proportional to the density of
harmonic measure $H(z)|\dd z|=|w'(z)||\dd z|$ of that point. As a
consequence, a sharper (more curved) part of the boundary moves with
a higher velocity than the rest, such that the sharper part is
getting even sharper, until it becomes singular. This occurs for 
generic initial data, although a number of exceptions are known
\cite{ms}. We will demonstrate this mechanism on a simple but
representative example, but first we remind the major feature
of the dynamics  \cite{Richardson72, PK}: let us denote by 
\be
\la{moments} t_k = -\frac{1}{k \pi}\int_{\tilde D} z^{-k} {\rm {d}}
x \dd y= \frac{1}{2 \pi \ii k }\oint_{\gamma}  z^{-k} \bar z\dd z,
\ee 
the harmonic moments of the fluid, and let the area of
the  domain $D$  not occupied by fluid  be $\pi t_0/\rho=Q\cdot t.$  Parameters $\{t_k\}$ are called {\it deformation parameters}. It follows from the Darcy law that \cite{Richardson72}
\be\la{Richardson}\dot t_k= 0.\ee

\begin{figure}[htbp]
\begin{center}
\includegraphics[width=7cm]{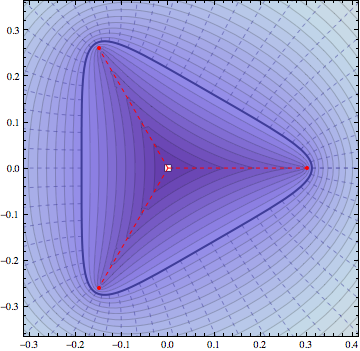}$\quad$
\includegraphics[width=7cm]{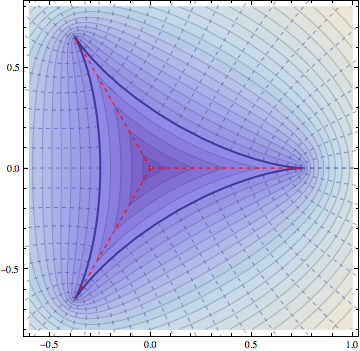}
\caption{Symmetric hypotrochoid  (\ref{hypo}) evolving under Darcy law (left) reaches  the (2,3)-cusp singularities  (right), when all three critical points (red dots) hit the boundary at the same time. The shaded contour lines are the equi-pressure lines. The dashed lines are the stream lines. The red dashed lines are  branch cuts -{\it a skeleton}.  }
\label{shypo}
\end{center}
\end{figure}

Let us write the inverse (time-dependent)  map  as \be \la{confmap}
z(w) = rw + \sum_{k \ge 1} u_k w^{-k}, \quad |w| \ge 1, \ee where
the coefficient $r$, chosen to be real positive is the conformal
radius (\ref{s}). According to a standard formula of complex analysis, the
coefficients of the inverse map determine the area of the domain $D$,
which we denote by $\pi t$, through  the {\it area formula} \cite{Ahlfors} 
\be \la{area} t = r^2 - \sum_{k
> 0} k |u_k|^2. \ee Let us consider a simple example of a domain $D$ 
whose shape is given by a symmetric hypotrochoid  - an algebraic curve with a
three-fold symmetry such that all harmonic moments vanish $t_k=0$,
except for the third one $t_3$ (see Figure~\ref{shypo}): 
\be \la{hypo} 
z(w) = rw + \frac{u}{w^2}. 
\ee 
The ratio between the square of  conformal radius
$r$ and $u$ can be expressed through $t_3$, as $u = 3\bar t_3 r^2$.
Then the area formula reads $t = r^2 - 2(3|t_3|)^2 r^4$.

Clearly, this polynomial in $r^2$ reaches the  maximum $t_c = r_c^2/2$ at $r_c=1/(6|t_3|)$,
and cannot increase further. At this moment, the hypotrochoid develops
three simultaneous cusp-like singularities, Figure~\ref{shypo}.

The mechanism of the evolution toward a critical point is: a
critical point of the  inverse map  $z'(w_c) = 0$ is  located outside of the fluid
 at $w_c=(2u/r)^{1/2}=(r(t)/r_c)^{1/2}<1$ and moves in time
towards the boundary.  At a critical time,  the critical point
reaches the boundary, at $w=1$. 

In the critical regime, when the
critical point of the conformal map is already close to the boundary
 $z(w)|_{w=e^{i\phi}}-\frac{3}{2r_c}=x(\phi)+ \ii y(\phi)$, the  boundary is 
 locally  approximated by a {\emph {real degenerate elliptic curve}}:
\bea\label{13}
 X(\phi):=\frac{2}{3}\frac{x}{r_c}\approx e(t)-\phi^2,&&\quad Y(\phi):=4\frac{y}{r_c}\approx 2\phi^3-3e(t)\phi,\\
  Y^2=-4 \left(X-e(t)\right)\left(X+\frac{e(t)}{2}\right)^2,&&\quad
e(t)=-\frac{4}{3}(1-\frac{r}{r_c})=-\frac{2}{3}\left(1-\frac{t}{t_c}\right)^{1/2}.
 \la{16}
 \eea 
At the critical point $t=t_c$, the curve degenerates further  to a  (2,3) cusp:
\be
Y^2 \sim X^3.
\ee
Close to the critical point, the conformal radius depends on time in a singular manner as $ r/r_c-1 \approx - \frac{1}{2}\left(1-t/t_c\right)^{1/2}$, so that the time dependence of power becomes non-analytic: $N-N_c\approx - \frac{1}{2}\left(1-t/t_c\right)^{1/2}$.

Once the (2,3) singularity occurs, the  shape of the pre-cusp finger
is given by the elliptic curve (\ref{16}). Up to a scaling of
coordinates and time the curve is universal - it does not depend on
details of the original domain:
\begin{align}\label{selfs1}
Y(X,t)=\left(1-\frac{t}{t_c}\right)^{3/4}Y\left(\frac{X}{(1-\frac{t}{t_c})^{1/2}}\right).
\end{align}
Self-similarity is preserved after the flow passes through the critical point, $ t>t_c$.

This simple analysis has a straightforward generalization for all
cases where a {\it finite} subset of $\{ t_k \}_{k > 0}$ is non-vanishing.
At least for this class of  initial conditions, the flow  leads to
cusp-like finite-time singularities. For purposes of analysis of
singularities it is sufficient to consider only this set of domains.
Of course, the class of initial conditions which leads to
singularities is much wider. We do not attempt to classify it here.
We only mention that   domains with a finite number of non-zero
exterior harmonic moments belong  to the class of {\it generalized
quadrature} domains. For generic properties of generalized
quadrature domains, see e.g., \cite{Bell04}.

\subsection{The first classification of singularities}

The fact that for rather general  initial data, the boundary evolves to a
cusp-like singularity had been noticed already in the earlier
papers \cite{Galin, PK} where the model of Laplacian Growth was originally 
formulated. Further developments are found in \cite{ST, bs84}, and  \cite{ Howison86, Howison85, Mineev90, Shraiman, Hohlov-Howison94,King}.  For the most generic initial conditions, a cusp (2,3) occurs.  Higher-order cusps and corner-like singularities require
special initial conditions.

In \cite{us4}, it was argued that, if a complement of a fluid domain  $D$ is a generalized
quadrature domain, then cusps of the type $(2,2l+1):\;
y^2 \sim x^{2l+1}$ are possible,  and  that in the
critical regime a finger developing into these cusps is described by a
{\it real degenerate hyperelliptic curve} 
\be\la{hp} Y^2
=-4 (X-e(t))\prod_{i=1}^{l} (X-d_i(t))^2.
\ee 
This means that a real oval of the curve - a set  where both  $X$ and $Y$ are real - is a graph of the finger $(x,y)=(X,Y)$.
The double roots $d_i$ ({\it double points}) are all located in the fluid domain and  simple
critical points  $e$ and $-\infty$ are located outside of the fluid. 

The condition that no critical points are found in the fluid, necessarily means that the
exterior critical points coincide, giving double points, such that
the complex curve is degenerate. A cusp of type $(2,2l+1)$ occurs
 when the  branch point  (real root $e$)  merges with $l$ double points.

For  reference,   we give  explicit formulas for the generic  hyperelliptic curve (\ref{hp}) following the Hele-Shaw evolution \cite{us4}. The curve depends on $l$ deformation parameters $\{ t_{2k+1}\}, k = 1, \ldots, l$ (not to be confused with the harmonic moments $t_k$), and it is uniformized by the formulas 
\bea\la{hpe}
&X=e(t)-\phi^2,\quad  Y=\sum_{n=1}^{l+1}(n+\frac12) t_{2n+1}\sum_{k=0}^{n-1}\,
\frac{(2n-1)!!}{(2n-2k-1)!!}
\frac{(-e(t))^k}{2^kk!}\phi^{2n-2k-1},\\
&  t_1+\sum_{k=1}^{l+1}\frac{(2k+1)!!}{2^kk!}t_{2k+1}(-e(t))^k = 0.
\eea
Here $t_1$ is   a  time before  the cusp singularity, $t_1\sim t-t_c$.

In \cite{Howison86} it has been argued that  dynamics can be
continued through a cusp of the type $(2, 4k+1)$ for the price of an appearance of an unstable point when a boundary touches itself.  Another stable 
solution  was  found in  \cite{us6} when the fluid undergoes a {\it genus transition}: it becomes multiply connected.  

However, in the case of the most generic cusp (2,3) and its descendants $(2, 4k-1), k > 0$, no smooth solutions are possible. This situation is the subject of this paper.

\section{Algebro-geometric formulation of Hele-Shaw problem}
\subsection{ Real algebraic curve}
  We briefly review some elements of the
 algebro-geometric description of the Hele-Shaw flow. 
 It is based on the notion of   real complex curve.

A boundary of an algebraic domain (the only case we consider)  is described by an equation $f(z,\bar z)=0$, where $f(z,\bar  z)=\sum_{nm}a_{mn}z^n\bar  z^m $  is a polynomial.  
   It is proved to be convenient to consider a Riemann surface or real algebraic curve defined by the equation $f(z,\tilde z)=\overline f(\tilde z,z)=0$,  where $\tilde z$ and $z$ are two complex variables. 
   For example, a curve for a symmetric hypotrochoid (\ref{shypo}) is $f(z,\tilde z)=(z\tilde z)^2-4r_c(z^3+\tilde z^3)+4(r_c^2-\frac{r^2}{4})(1+\frac{r^2}{2r_c^2})(z\tilde z)-4r^2(r_c^2-\frac{r^2}{3})$ \cite{us9}.

Among many solutions of the equation $f(z,\tilde z)=0$  with respect to $\tilde z$ (sheets of a  curve), there is one $\tilde z=S(z)$ (a physical sheet) which  describes a boundary where $\tilde z=\bar z$,
\be \la{S} 
S(z) =  \bar z, \quad z \in \gamma. 
\ee

 It follows from Darcy law  that the time derivative of   function $S(z)$  is
holomorphic in the fluid. On the boundary and in the fluid domain $\widetilde D$ a complex velocity $v=v_x- \ii v_y$ is obtained by $2v=\p_t \bar z=\p_tS(z)$.
Everywhere in the fluid the velocity  is holomorphic. In terms of the function $S$
 the Darcy law reads:
\begin{align} \la{darcy20}
2v(z,t)=& \p_t S(z, t),\\
\la{darcy2} \p_t S(z, t)= & 2 \ii \p_z \phi(z,t), 
\end{align}
where 
\be \la{potential}\phi(z,t) = \psi(z,t) + \ii p(z,t) \ee
 is the potential of the flow, whose
imaginary part is the usual scalar pressure $p$, and the real part is the stream function
$\psi$. Wherever  pressure is harmonic, the potential is an analytic function.

 As we have said,
   to study local behavior of  an isolated cusp, it is sufficient
   to consider domains with a finite set of non-zero harmonic moments (\ref{moments}).
In this case the function $S$ has a multiple pole at infinity. It 
 is represented by a truncated Laurent series with respect to a drain at infinity and an arbitrary regular  point, say $z=0$, outside of  the fluid:
\be \la{expansion} S(z) =S_+(z)  +  S_-(z),\quad S_-(z)=tz^{-1}+\sum_{k>1}^\infty v_k z^{-k},
\ee
where we set the  positive part to be a  polynomial: 
 \be \la{22}
S_+(z)= \sum_{k > 0}k t_k z^{k-1}. 
\ee
It  encodes information about the  moments of the exterior \cite{S3} and from (\ref{darcy20}) it does not evolve 
in time (\ref{Richardson}):
\be
\dot S_+(z,t)=0.
\ee
Parameters of the flow -- time $t$  and deformation parameters $t_k$ -- can be seen as residues of the differential $ z^{-k}S(z)\dd z$ at $z=\infty$ (a drain).

{\em Before a critical time},  complex velocity of the  fluid is  holomorphic everywhere in the fluid domain. In this case,  the negative part  of  the function $S(z)$  is an analytic continuation  of $\bar z$ to the fluid domain:
\be\la{S5}
S_-|_{z\in \tilde D}=\frac{1}{2\pi\ii}\oint_{\gamma}\frac{\bar\zeta\dd\zeta}{z-\zeta}=\frac{1}{\pi}\int_{D}\frac{\dd^2\zeta}{z-\zeta}.
\ee
In this case, $S(z)$ is called a Schwarz function \cite{Davis}. All singularities of $S_-(z)$ are branch points located outside of the fluid domain.

A generating function  -- integral of the meromorphic differential $\dd\Omega=S(z)\dd z$ from a point of the boundary of the fluid $z_\gamma$ to a point of the fluid
\be\la{Omega5}
\Omega(z)=\int ^z_{z_\gamma}S(z')\dd z',
\ee
is a convenient way to describe Hele-Shaw flows. For example,  the boundary of the fluid  given by  (\ref{S}) is a solution of the equation
\be\la{Omega6}
|z|^2=2{\rm\,Re}\,\Omega(z).
\ee
A graph of $-|z|^2+2{\rm\,Re }\, \Omega(z)$ close to a critical point is plotted  in Figure~\ref{OmegaFIG}.
\begin{figure}[htbp]
\begin{center}
\includegraphics[width=8cm]{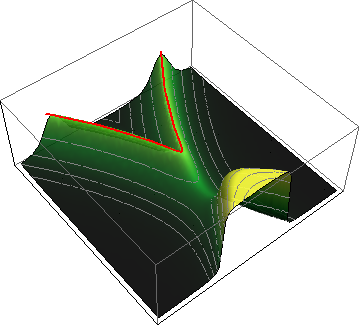}\quad
\includegraphics[width=8cm]{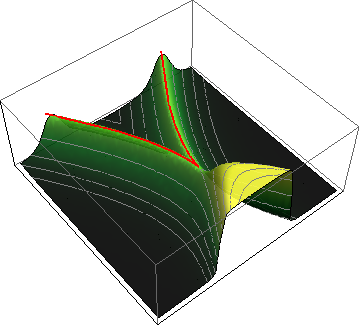}
\caption{The plot of $-|z|^2+2{\rm Re}\int S(z) \dd z$ before a critical time  (left) and at a critical time (right).  The height difference has been blown-up by scaling the vertical axis and by applying $\arctan$. \la{OmegaFIG}}
\end{center}
\end{figure}

In a critical regime, i.e., close to a cusp, it is convenient to describe the boundary  in Cartesian coordinates $Y=\frac{1}{2\ii}(z-S(z))$ and $X=\frac{1}{2}( z+S(z))$,  \la{R2} and treat $Y$ as a function of $X$. In these coordinates  \cite{us4}
   Darcy law (\ref{darcy20}) reads: \la{R3}
 \be \la{darcy3}
\p_t Y(X, t)=  -\p_X \phi(X,t). 
 \ee
This follows from an important  property of the Darcy law (\ref{darcy20}):  the form $\dd\omega=S \dd z + 2\ii \phi \dd t $ is closed. 
Therefore, the Darcy law is invariant under canonical transformations $(z,\, S)\to (X,\,Y):\;\:\dd\omega=-2\ii Y\dd X + 2\ii \phi \dd t $, up to an exact form.

 In a critical regime,  it is also convenient to choose a critical point  as origin and to redefine  time $t\to t_1\sim t-t_c$ and deformation parameters as $t_{2k+1}=\frac{1}{2(2k+1)}{\rm res}_{\infty}(-X)^{-k-1/2} Y\dd X$.   
Then, in the case of  a $(2,2l+1)$ - cusp,  a hyperelliptic curve (\ref{hp}) is a Laurent series with respect to the local parameter at infinity $X^{1/2}$:
 \bea\la{YY}
&Y(X)=Y_+(X)+Y_-(X),\\
& \la{Y7} Y_+(X)=\sum_{k=0}^{l+1} (k+\frac12)t_{2k+1}(-X)^{k-1/2},\\ 
&\la{C0} \phi(X)=-( -X)^{1/2}+\frac{C}{12} (-X)^{-1/2}+\phi_-(X),
\eea
where $Y_-$ and $\phi_-$ consist of further negative powers in $X^{1/2}$. Here, $C$ is the capacity of a  finger -- relative to the capacity at a critical time (capacity for a compact set is defined just  below \eqref{s}). Darcy law then states that the positive part is conserved:
\be\la{YY}
\dot Y_+(X)=0.
\ee
Eqs. (\ref{hp}) give solutions of the Darcy law written in the form (\ref{darcy3}) under assumption (\ref{Y7}). The parameter $\phi$ in (\ref{hpe})  (treated as a function of $X$) is the complex potential of the flow at the point $z=X + \ii Y(X)$.

\subsection{Skeleton and inverse {\it balayage}}\la{skeleton}
If the number of non-zero harmonic moments is finite,
 the  function $S(z)$ (\ref{expansion})  has only a finite-order multiple
 pole  at infinity  and branch points in the finite part of the complex plane. 
 Before a critical point forms, all singularities are moving branch points located 
 outside of the fluid. Their dynamics encode the entire flow.

\begin{figure}[htbp]
\begin{center}
\resizebox{0.45 \textwidth}{!}{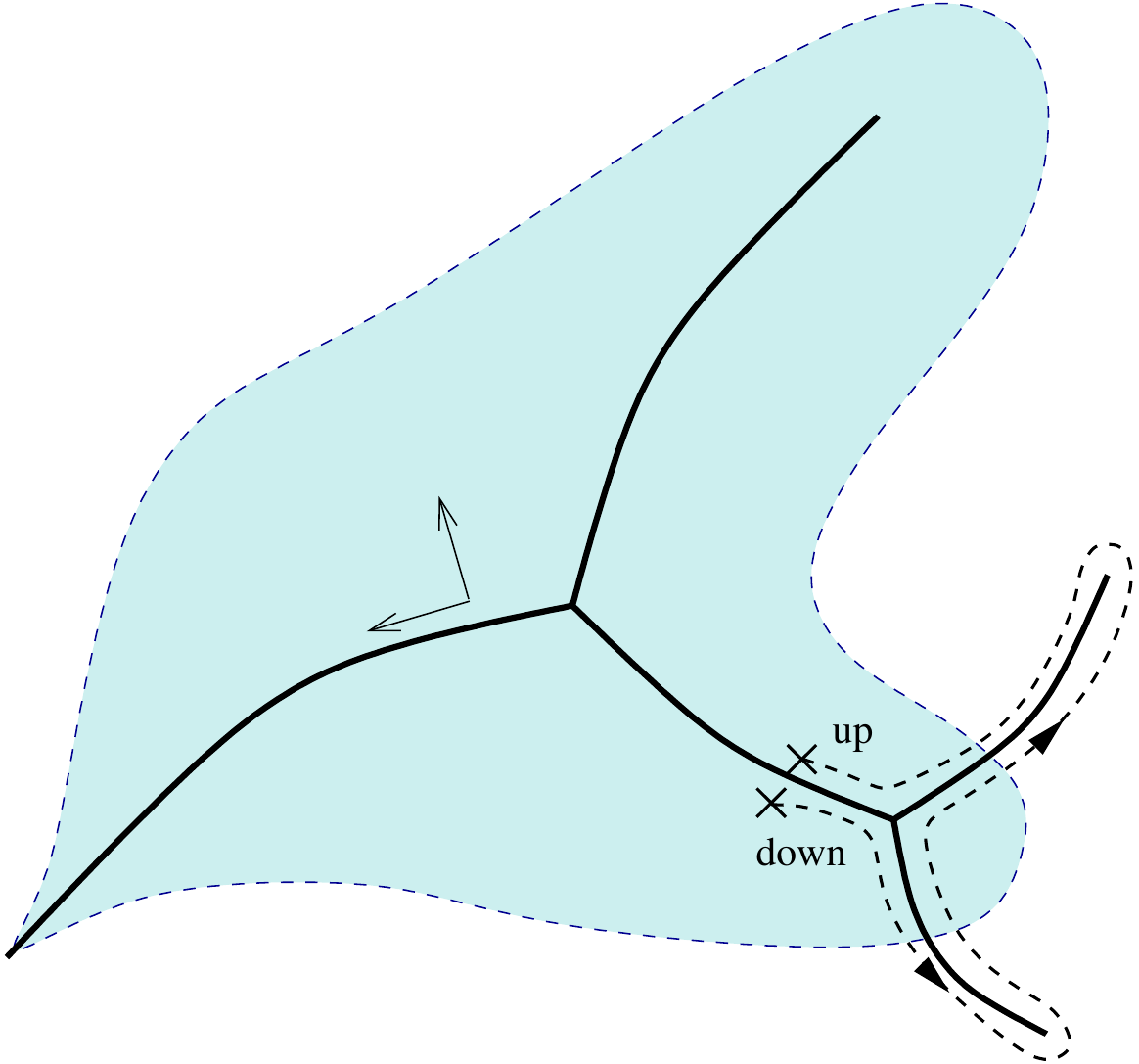}
\resizebox{0.45 \textwidth}{!}{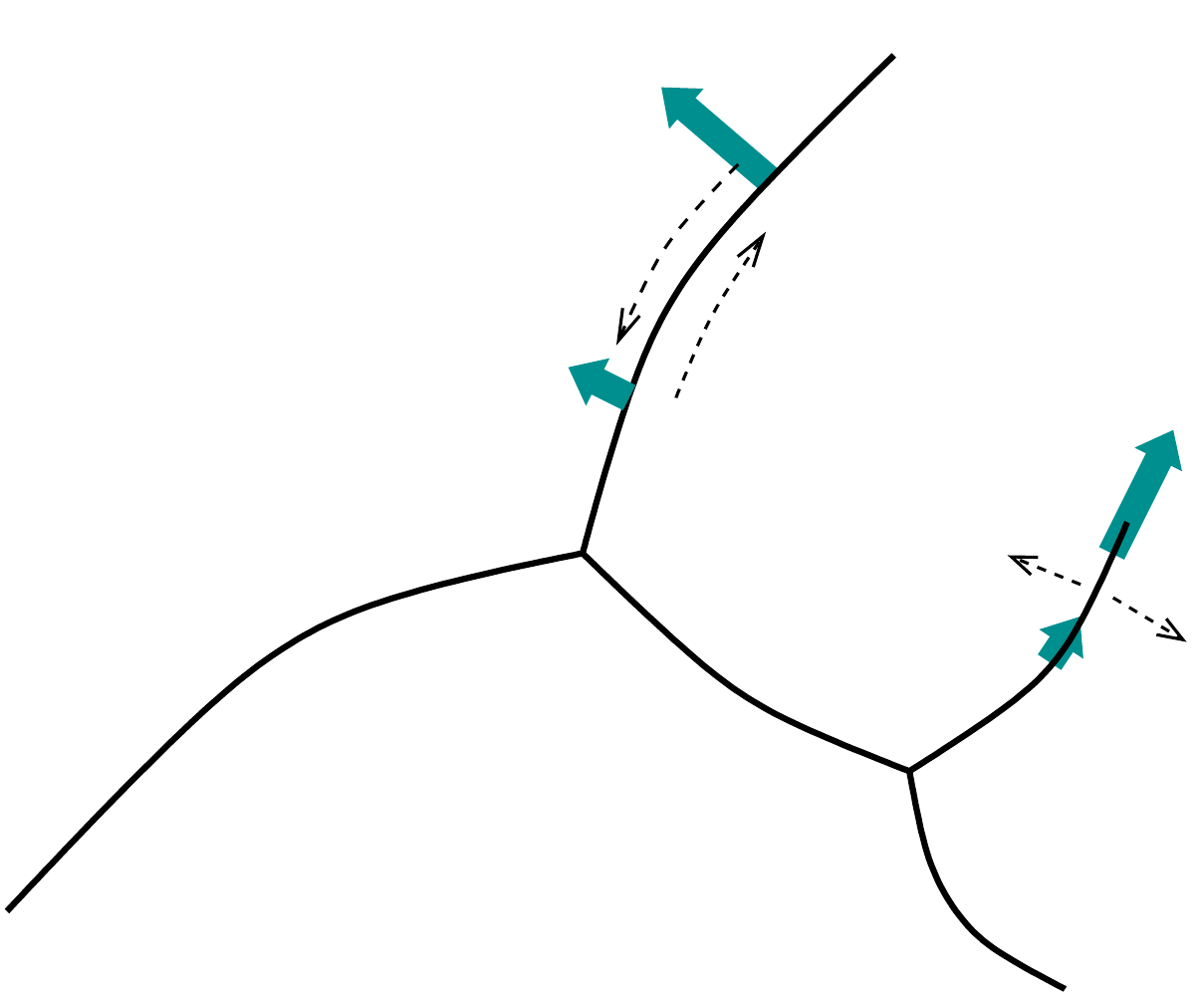}
\caption{
In the left panel, we show the three possible stages of evolution: i) when a branch point is outside the fluid, ii) when a finger forms a cusp (a branch point is on the boundary), and iii) when a shock develops after a cusp (branch points are inside the fluid).   Canonical orientation of the contour (dashed line $\Gamma$)  define how the discontinuities are taken.    In the right panel, we show the normal velocity of the shock by the thick arrows and the flow of fluid around the shock by dashed arrows.  The arrows illustrate that  vorticity carried by  a shock is compensated by the circulation of the fluid around the shock.  }
\label{orientation} 
\end{center}
\end{figure}

In this case, there is a unique, special way to draw  branch cuts.
They can be  drawn as a (possibly multi-component) curved graph 
$\Sigma(t)$ such that  the  jump of the differential $-\ii \dd\Omega = -\ii S \dd z$, 
being canonically (counterclockwise) oriented along the curve through 
every branch cut, will be real and positive: 
\be\la{jump} 
-\ii\, {\rm{disc}}\; S(z,t)\dd z=2\sigma_k (z,t)|\dd z
|>0,\,\, z\in \Sigma_k (t),
\ee 
where $k$ labels branches of the graph.
Here the differential is taken
along the cut and $|\dd z|$ is an element of an arc-length along a cut. 
The canonical orientation (important for all signs in formulas below) is depicted in Figure~\ref{orientation}.

If $S$ is treated as a complex vector, then (\ref{jump}) reads
\be\la{jump'} 
\overline{ {\rm{disc}}\; S}=2\sigma_k {\bf n}.
\ee

This fact is equivalent with the following analytical continuation 
of $S_-(z)$  (\ref{S5})  to the domain $D$:
\be \la{integ}
 S_-(z) =\frac{\rho_0}{\pi}\int_{D}\frac{\dd^2\zeta}{z-\zeta}= \sum_k \int_{\Sigma_k} \frac{\sigma _k(\zeta)}{z - \zeta} |\dd \zeta|,
\ee
with $|\dd \zeta|$ the arclength measure. In other words, $S_-(z)$  is 
the electric field produced by a single layer of charges $\sigma _k$ distributed along a graph $\Sigma$.  An important theorem on generalized 
quadrature domains \cite{G4} tells that $\sigma $ exists and  is unique. The curved and
generally multi-component graph $ \Sigma$, 
confined in the domain $D$, is called {\it skeleton} or {\it mother-body} 
\cite{G4}.  Line densities of the skeleton depend on time only through time
dependence of the branch points. If a branch point $e(t)$ is
simple (which we assume), the line density $\sigma (z,t)$
vanishes at the branch point by a {\it semicircle law} as 
\be
\sigma (z,t)\sim \sqrt {|z-e(t)|} \ .
\ee
In the case of the hypotrochoid (\ref{hypo}), the graph consists of three curved lines
connecting the branch  points. They are drawn as in Figure~\ref{shypo}. In the case of symmetric hypotrochoid \eqref{hypo}, the branch cuts are three straight segments, connecting the branch points
$e(t)=\frac{3}{2}r_c\left(r(t)/r_c\right)^{4/3}\times (1,e^{\pm
i2\pi/3})$, with line densities at the branch points 
$4(2/3)^{1/2}r_c(1-r^2/r_c^2)(r_c/r)^{2/3}[\sqrt{|z-e(t)|}+(4/3)(r_c/r)^{4/3}|z|/r_c\sqrt{|z-e(t)|}]$ (where $z$ is measured from a  position of a cusp). \la{R4}

The major property   of generalized quadrature domains (actually
serving as definition) \cite{Bell04} is:  area averaging of an
integrable  analytic function $f(z)$ over any domain $B$
  is reduced to  counting of singularities of the   function $S(z)$
$\frac{1}{\pi}\int_B f(z) \dd ^2z = \frac{1}{2\pi \ii}\oint_{\p B}
f(z) S(z) \dd z.$ This follows from 
the representation (\ref{integ}):
\be \la{gqd}
\begin{array}{lcl}
\frac{1}{\pi}\int_B f(z) \rho\,\dd x \dd y & = &  \frac{1}{2\pi \ii}\oint_{\p B} f(z) S(z) \dd z \\
&& \\
& = &  \int_{\{ \Sigma\} \cap B} \sigma (z) f(z) |\dd z|.
\end{array}
\ee
If the domain
$B$ does not contain moving singularities or the drain, then the
average stays constant in Hele-Shaw flow.
Conversely, if the domain $B$ contains  moving singularities, or a
drain, the average depends on time.

The procedure of sweeping a domain $B$ into a line graph $\{\Sigma\}\cap B$ with a line density $\sigma $ is known as (inverse) {\it
balayage} \cite{G4}.  A Newton potential created by uniformly
distributed mass outside of the fluid, measured inside the
fluid, in $\tilde D$, will be the same as the Newton potential
created by the (non-uniform)  line densities $\sigma $.

The balayage procedure allows the following insightful
interpretation of Hele-Shaw flow. The  ``skeleton"  -- moving branch
cuts -- may be considered as  time-dependent sources of the viscous
fluid. In this interpretation, the fluid has no boundary,
only   time-dependent line sources. An extended fluid occupies all
the plane except the sources and moves  with velocity $v=(1/2)\p_t
S$. The pressure of the extended fluid  is defined by
$ v=-\nabla p$ . The original boundary  can then be restored (if
necessary) as a {\it real oval} of the complex curve - a planar
curve where $S(z)=\bar z$, or (equivalently) a zero-level set of the
pressure $p=0$. Darcy Law (\ref{darcy20}, \ref{darcy2})  connecting velocity and a complex potential then applies
both inside and outside the fluid.

In this interpretation, the entire flow is encoded by the motion of
the skeleton, or rather by the motion of end points of the skeleton.
Velocity, pressure and stream function generally  have finite discontinuities
across the skeleton. Also, pressure may not be constant along
the sides of the skeleton, while velocity is not normal to the
skeleton. 
A singularity occurs when a growing skeleton intersects the boundary of the fluid.

In the next section,  we formulate the weak solution of the Hele-Shaw flow and  derive  
equations for the moving shock graph. The same equations are applied for a moving skeleton. 
Vice-versa, an evolution through a cusp singularity will be ``continuous" only if shocks are governed  by the same law as the skeleton.

\section{Weak solutions of Hele-Shaw problem} \la{section_three}

\subsection{Weak form of Darcy law:  the Rankine-Hugoniot and admissibility conditions} 
Now we assume that after a singularity has occurred, pressure is  a harmonic function everywhere except on a moving  shock graph $\Gamma(t)\subset \tilde D$ located in the fluid domain, where pressure and stream function have finite discontinuities. Then so does the r.h.s. in  (\ref{darcy2}). 

The function $S(z)=S_+(z)+S_-(z)$ that we defined before the cusp occurs changes its property and 
is no longer a Schwarz function. In particular, it exhibits a discontinuity in $S_-(z)$ along the shock (in addition to the skeleton) such that:
\be\la{jump1} 
-\ii\, {\rm{disc}}\; S(z,t)\dd z=2\sigma_k (z,t)|\dd z
|>0,\quad z\in \Gamma_k (t)\ .
\ee 
We keep the same notations for line densities along shocks as lines densities along a skeleton (\ref{jump}).

We also assume that shocks that appear in the  post-critical regime do not affect the fluid far away from the critical point.
We require that  the lines densities $\sigma_k$ are {\it real} and {\it positive}, as  for a skeleton (Sec.~\ref{skeleton}). This is the major condition. We illustrate its meaning below.

Therefore, after the critical time, $S(z)$ develops branch cuts  in the fluid domain $\tilde D$ in addition to the branch cuts in $D$.  We write:
\be\la{sigma1}
S_-(z) =\frac{\rho_0}{\pi}\int_{D}\frac{\,\dd^2\zeta}{z-\zeta}+\sum_k \int_{\Gamma_k} \frac{\sigma _k(\zeta)}{z - \zeta} |\dd \zeta|=\int_{\Sigma \cup\Gamma} \frac{\sigma _k(\zeta)}{z - \zeta} |\dd \zeta|, \quad  {z\in \tilde D}.
\ee
With an assumed orientation as in Figure~\ref{orientation}, we can interpret (\ref{sigma1}) as a {\emph{deficit}} of the fluid density: 
$\rho=\rho_0-\delta(z;\Gamma)\sigma(z)$.

A comment is in order. Before a critical time, the function $S(z,t)$ in (\ref{jump1})  was analytic in a vicinity of the fluid boundary, and therefore was a Schwarz function of that boundary. Not so  anymore after a critical point. It jumps at branch cuts that protrude from a skeleton.
For  the same reason, a domain $D$ (complement to the fluid domain which includes shocks) is not a a generalized quadrature domain any longer.

 Let us 
integrate (\ref{darcy2}) over a closed loop  $\p B$ bounding a domain $B$ drawn
anywhere in the fluid. If the loop does not  cross a shock, we obtain
\bea \la{im} \frac{1}{2} \frac{\dd}{\dd t}{\rm{Im}}\,
\oint_{\p B}   S(z) \dd z&=&-  
\int_B \left(\mathbf{\nabla \cdot j}\right) \, \dd^2z =\oint_{\p B}\mathbf{ j\times dl}=\oint \dd\psi,\\
\la{re} \frac{1}{2} \frac{\dd}{\dd t}{\rm{Re}}\, \oint_{\p B}   S(z) \dd z
&=&
 \int_B  \,\left(\mathbf{
\nabla\times j}\right) \,\dd^2z =\oint_{\p
B}\mathbf{ j \cdot dl}=\oint \dd p, 
\eea 
where $\mathbf{
\nabla\times j}=\p_y j_x - \p_x j_y$ is the vorticity field, and
$\mathbf{ \nabla\cdot  j } = \p_x j_x + \p_y j_y$ is the divergence
of velocity field ($2\bar\p j=\mathbf{ \nabla\cdot  j } +\ii \mathbf{
\nabla\times j}$), and $\bf j=\rho v$.
The first two equalities in each line of (\ref{im}, \ref{re}) are
identities, the last is one is the Darcy equation.

Let us now consider the integral 
$\oint_{\p B} S(z) \dd z$ over a loop which does intersect 
a shock graph at one or two points. This integral measures the density of the portion of a shock surrounded by  the loop and under  (\ref{jump1}) stays
 purely imaginary at all times. We conclude that
\be \la{krichever}
\frac{\dd \,\,}{\dd t} {\rm{ Re }} \oint S(z) \dd z = 0. 
\ee
Letting the contour now shrink to an infinitesimal loop, we obtain the
differential form of (\ref{krichever}) as:
\be \la{diff_krichever}
\frac{\dd \,\,}{\dd t}\,\, {\rm{Re }} \, [{\rm{disc }} \, S \,\, \dd z ]_{\Gamma} = 0. 
\ee
The total time derivative in (\ref{diff_krichever}) has  two
contributions: one from the time evolution of the  function $S(z)$, the other
from the motion of shocks. We denote the  velocity of the 
shock front (normal to the instantaneous curves $\Gamma$), 
by ${\bf V}_\perp$, directed along the vector $\bf n$ as in Figure~\ref{orientation}. Then the total time derivative (\ref{diff_krichever}) becomes
\be \la{ident}
{\rm{Re }} \, [{\rm{disc }} \, \dot S \,\, \dd z +\nabla_\parallel ({\rm{disc }} \, S \cdot {\bf V}_\perp) |\dd z| ]_{ \Gamma}
=0,
\ee
where $\nabla_\parallel$ represents the derivative along the direction tangent to the front along the vector $\bm\ell$
as in Figure~\ref{orientation}. 

For the first term, we use a kinematic identity 
${\rm{Re }} \, [{\rm{disc }} \, \dot S \,\, \dd z] = 2 {\rm{disc}} \,\, v_{\parallel} |\dd z|$, valid on both sides of the shock.
From the reality of the jump (\ref{jump1}) it follows that the second term in (\ref{ident}) is purely 
real, and equals $2 \nabla_\parallel \left(\sigma {\bf V}_\perp\right) |\dd z|$. 

Together, it yields to the condition
\be\la{C}
\nabla_\parallel{J_\perp} + \,{\rm{disc}} \,\, j_{\parallel} = 0,\quad {J_\perp}= \sigma  { {\bf V}_\perp},\quad j_\parallel=\rho_0 \,v_\parallel.
\ee
The first term in this equation represents the transport of mass due to motion of the shock (normal to the shock itself), while the
second is the vorticity of the surrounding fluid flow. They compensate each other. This condition, derived solely from the requirement that $\sigma $ is real, suggests to interpret a  shock   as a single layer of positive vortices. Then $\sigma $ is the (smooth part of) density of vortices. The core of the vortex is of the order of $\hbar^{1/2}$ -- the width of the shock. 
Then the reality condition for $\sigma$ means  a {\it zero-vorticity} condition for the fluid.

Using Darcy law we  replace the fluid velocity in (\ref{C}) by
$-\nabla_{\parallel} p$, and integrate (\ref{C}) along the cut. We obtain the  Rankine-Hugoniot condition
\be\la{rh1}
 \sigma  {\bf {V_\perp}}=({\rm {disc}}\;p)\;{\bf n}.
 \ee
(the constant of integration is fixed from the assumption that both the line density  and  discontinuity of pressure vanish at endpoints of the cut). 
The fact that $\sigma>0$ is positive means that a shock moves toward the direction of larger pressure.

Furthermore, calculating the discontinuities on both sides of the Darcy law and using (\ref{jump1}) we obtain
\be\la{dot}
\dot \sigma  =-\rm {disc}\;\nabla_\parallel\psi\ .
\ee
This formula has the following interpretation. Let us assume the shock  filled with some material (say, an inviscid fluid in the Hele-Shaw cell), of line density $\sigma$. Then 
the continuity condition $\dot \sigma +\nabla_\parallel (\sigma  {\bf V}_\parallel)=0$,  tells that this material is moving along the shock with velocity ${\bf V}_\parallel$. Combining this with (\ref{dot}) and integrating along the shock,  we obtain a counterpart of (\ref{C}):  divergency of the sliding current along a shock  equals the discontinuity of current of fluid normal to the shock (see Figure \ref{orientation}):
\be\la{d}
\nabla_\parallel{J_\parallel} -{\rm{disc}} \,\, j_{\perp} = 0,\quad {J_\parallel}= \sigma  { {\bf V}_\parallel},\quad j_\perp=\rho_0 v_\perp.
\ee
Integrating this formula along a shock gives a counterpart of the  Rankine-Hugoniot condition:
\be\la{rh2}
 \sigma  \bf{V_\parallel}= (\rm {disc}\,\psi)\: \bm \ell.
 \ee
 At an end  point of the shock $e(t)$,  the total velocity of the shock matches the velocity of the branch point:
 \be
 \dot e(t)=\left({\bf V}_\perp+{\bf V}_\parallel\right)|_{z=e(t)}.
 \ee
 We notice that  (at least close to the endpoint),  the stream of mass  is directed towards the branch point.

The Rankine-Hugoniot conditions (\ref{rh1}, \ref{rh2}), inequality  (\ref{jump1})  and  the differential  form of the Darcy equation (\ref{darcy2}) combined give the weak forms of the Darcy law. 

 The same conditions hold for a  skeleton before and after critical time.  After  a critical time,  shocks can be considered as continuation of growth of a skeleton inside the fluid. From this point of view, skeleton and shocks together form a graph which grows under the same law. As we will see below, the boundary is singled out as a position
of the first  branching event. Following branching events already occur in the fluid as branching shocks. 
 
Summing up: three conditions
\be\la{weak}
 \sigma {\bf{ V_\perp}}=({\rm {disc}}\;p)\;{\bf n},\quad \dot \sigma  =-\rm {disc}\;\nabla_\parallel\psi,\quad \sigma>0,
 \ee
together with  the differential Darcy law outside of the shock's graph constitute {\it the weak form of Darcy law}.

We note that this weak solution only applies to {\em algebraic solutions} unlike the conventional Hele-Shaw problem. Though condition \eqref{weak} tells one how the  shocks and domains evolve in the next moment, the determination of the driving force---pressure---is not given by the simple Dirichlet problem as in the conventional Hele-Shaw problem.  To obtain pressure one needs to know $S(z)$,  which contains a finite number of branch cuts and poles. If the  function $S(z)$ is algebraic  at some moment in time, integrability of the problem guarantees that it will remain so at all times.  

\subsection{Algebro-geometrical formulation of the weak solution}\la{real curve}

Having  real complex curve in mind (not only its real oval) has proved to be useful for studying a Hele-Shaw flow and is necessary to understand its weak solutions.   A Hele-Shaw flow of  an algebraic domain is then understood as  the  evolution of  a meromorphic differential $\dd\Omega=S(z)\dd z$. The properties described above can be cast as algebro-geometrical properties of an algebraic curve:
\begin{enumerate}
\item \la{i1}  the complex curve is real;
\item the  condition that all densities  are real yields
\be \la{boutroux}
{\rm{Re }} \,\,  \oint \dd \Omega = 0,\quad {\rm{for\; all\; cycles}~\mbox{in the fluid;}}
\ee
\item\la{SS}
shocks form a graph determined by the condition
\be\la{A1}
{\rm Re}\,{\rm{disc}}\,\Omega=0,
\ee
and the admissibility condition \ref{31};
\item \la{31} the condition that all densities are positive yields
that  ${\rm Re}\,\Omega$ is {\it increasing} away from the cut; 
\item the meromorphic differential $\dd\Omega$ has multiple poles at infinity, the residues at infinity of the differential $z^{-k}\dd\Omega$ are the deformation parameters (\ref{moments}), and time is the residue of the differential  $\dd\Omega$ at infinity.
\end{enumerate}
The graph (\ref{A1}) consists of curved lines, sometimes called {\it anti-Stokes} lines. Condition 3 selects {\it admissible}  anti-Stokes lines.

Complex curves obeying condition  $2$ are called  {Krichever-Boutroux} curves.   Evolution of a curve obeying the Krichever-Boutroux condition through a change of genus   had already
  appeared in studies of Whitham averaging in integrable systems by Krichever \cite{Krichev-red}.
Similar condition has been introduced by David  in the theory  of 2D quantum gravity \cite{David},  and  recently was recognized in studies of distribution of zeros in orthogonal polynomials by 
Bertola and Mo \cite{Bertola-Mo}. Neither appearance is a  coincidence.  Semiclassical asymptotes of orthogonal polynomials and Whitham genus changing transition 
are ultimately related to the Hele-Shaw flow.

According to property $3$, the real part of the {\it generating function} $\Omega(z)=\int^z_{z_\gamma} S(\zeta)\dd\zeta$, where $z_\gamma$ is a chosen point on the boundary, is a single-valued function.  The integral of the differential over a cycle involving a drain is $\oint\dd\Omega=2\ii  t$. 

Being  cast in this form the properties of the curve  fully and uniquely determine a smooth  Hele-Shaw solution before a singularity is reached, and a weak dispersive solution afterwards.

As we have discussed, the boundary of the fluid (before it reaches a singularity) is a degenerate curve. A singularity occurs when a branch  point meets a double point on a boundary, forming a triple (or higher order)  point, so the curve  further degenerates, while the boundary of the real oval forms a cusp. Beyond this time, the triply-degenerate  point  splits into three regular branch  points, so that the curve becomes non-degenerate. In other words, going through a singularity, the curve changes genus. This process is illustrated in Figure~\ref{beforeafter}.

In a critical regime (i.e., close to a cusp-singularity), a complex
curve locally is   hyperelliptic $Y^2=R_{2l+1}(X)$ (\ref{hpe}) and 
\be
\Omega(X)=- \ii \int^X_{-\infty} Y\dd X,
\ee
where $-\infty$ is a distant point on a skeleton (negative real axis).
In this case  ${\rm Im}\;\Omega$ on both sides of the cut has opposite signs, while ${\rm Re}\,\Omega$ is zero on the cut. For hyperelliptic curves, conditions 3  and 4 are formulated in terms of level lines of the field of the potential ${\rm Re}\,\Omega$:
\begin{align}
\la{AA}
{\rm{Re}} \,\Omega (X) > 0, \quad X\to \Gamma;\\
\label{A2} 
{\rm Re}\,\Omega(X)|_\Gamma=0 \ .
\end{align}
Shocks and skeleton are zero level lines of the potential ${\rm Re}\,\Omega$, such that the potential stays positive in the neighborhood. Then the line density of shocks and the mass deficit accumulated by a shock between two points $X_1$ and $X_2$ are:
\begin{align}\la{Y}
\sigma(X)=|Y(X)|_\Gamma,\\
\la{n}\int^{X_2}_{X_1}\sigma |\dd X|=|{\rm Im}\,\Omega(X)|\Big|^{X_2}_{X_1}.
\end{align}
 In the next section. we show how these two conditions uniquely determine the evolution of the elliptic curve through the critical time.

\section{Beyond the (2,3)-cusp singularity: self-similar weak solution} 

Here we give a detailed analysis of the generic (2,3)- cusp singularity and illustrate the nature of weak solution. The computations below describe the evolution of a unique elliptic Krichever-Boutroux curve.  We emphasize that part of this analysis appeared
in the papers  \cite{David,Gamba, Novikov}, which is devoted to  Whitham averaging of the 
triply-truncated  solution  of Painlev\'e I equation.
In this section we again  set $\rho_0=1$.

We study the evolution of the elliptic curve (\ref{13}-\ref{selfs1})
\be\la{133}
Y^2=-4 \left(X-e(t)\right)\left(X+\frac{e(t)}{2}\right)^2,
\ee
 through the critical point where $e(t)|_{t=0}=0$. It is in order to remind  how complex coordinates $X$ and $Y$ are related to a coordinate $z=x+ \ii y$ of the fluid. It is $z/r_c=X+ \ii Y(X)$. On the boundary of the fluid $X$ and $Y$ are real and they  coincide  with Cartesian coordinates $(x,y)$ of the fluid. Elsewhere $X$ is complex. The coordinate of the fluid is given by the map $X\to z$. The map covers the physical plane twice, as shown in Figure~\ref{rhombuspressure}, so one must choose a physical branch of the map. Under this  map a point  in the fluid $z$ is in one-to-one correspondence with a point on the $X$-plane, except on the fluid boundary. There, two points $z$ and $\bar z$ correspond to the same $X$.  As it follows from (\ref{13}) $X$ and $Y(X)$ have different scaling $X\sim x/r_c,\;Y\sim (x/r_c)^{3/2}$.  
 Therefore at $x\ll r_c$ i.e., at the critical regime where approximation of the finger by the elliptic curve make sense, $Y(X)\ll X$ and therefore $z\approx X$.  Having this fact in mind  below  we express the flow in the $X$-plane. There the boundary of the fluid -- a narrow finger - is represented by its skeleton
 -- a cut on the negative axis ${\rm Im}\, X=0,\,{\rm Re}\, X<e=e_3$, as in Figure~\ref{beforeafter}.

\subsection{Elliptic curve -- genus transition}

\begin{figure}[htbp]
\begin{center}
\includegraphics[width=6cm]{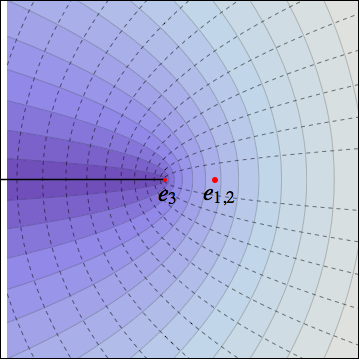}
\includegraphics[width=6cm]{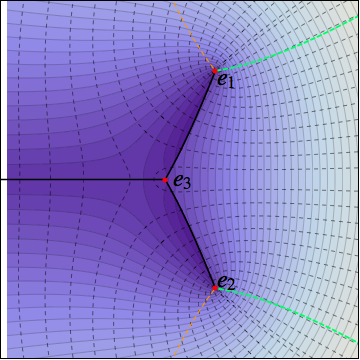}
\caption{The equi-pressure lines and the flow (dashed) lines before the  
critical time (left), and after the critical time  (right).    Pressure of fluid gets larger as the shade gets  
darker.  The boundary of the fluid is a very narrow finger $y^2\sim  
x^3$ around the skeleton line (the  thick line on the  real axis) and  
it is not depicted. The red dots are the branch points, $e_{1,2,3}$.  
Real $e_3$ is the tip of the finger. Before the critical time  
$e_2=e_1$ is the double point (larger red dot in left figure). The  
thick lines (right panel) connecting the branch points  are  
the shocks. They emanate with the angle $2\pi/3$ and diverge at their  
ends by a slightly large angle $0.0918498\pi+\frac{2\pi}{3}$. In the  
 figure on the right, the orange and green dashed lines that emanate from the  
branch points are not-admissible  level lines of ${\rm Re}\,\Omega$.
The fluid flows to the lighter  region, toward low pressure. Shocks move toward the darker region (higher pressure).
}\label{beforeafter}
\end{center}
\end{figure}

Figure~\ref{beforeafter} illustrates the genus transition of the curve and hydrodynamics of shocks. 
Before the critical time, two branch points coincide in the fluid. There, pressure and velocity are smooth. The branch point $e_3$ is the tip of the finger. At the critical time,  the tip meets the double point. After the critical time,  the double point splits into two branch points.  They are the endpoints of the shocks. The tip of the finger $e_3$ retreats.  Fluid goes towards lower pressure (lighter regions), creating a deficit  on shocks where fluid decompresses. Shocks move in the opposite direction,  towards higher pressure (darker regions).

We write the curve as:
\begin{equation}\la{58}
Y^2=-(4X^3-g_2(t)X-g_3(t))=-4(X-e_3(t))(X-e_2(t))(X-e_1(t)).
\end{equation}
where  $e_3(t),\,e_2(t),\,e_1(t)$ are time-dependent branch  points. One root, say $e_3$, can always be chosen to be real. The other two  are complex conjugated $e_2=\bar e_1$. The coefficient in front of $X^2$ does not depend on time and  can be removed by translation, such that  $e_3+e_2+e_1=0$. 

The positive part of the curve is  $\pm Y_+=2 \ii X^{3/2}-\ii \frac{g_2}{4}X^{-1/2}$.  
The coefficient of the $X^{-1/2}$-term is is proportional to time. We normalize it as
$-\frac{1}{4}g_2=3t$ (this choice corresponds to  $t_1=6 t$ in \eqref{Y7}, which further requires a change $\phi\to \phi/6$ below). 
Under this setting,
\be\la{g}
-\frac{1}{4}g_2=-e_3^2+|e_2|^2=3t,\quad g_3=4e_3|e_2|^2,\quad e_3=-2\mbox{Re}\,(e_2).
\ee
\begin{enumerate}
\item 
Before the critical time $(t<0)$ the real branch point $e_3$ is negative. It is located on the boundary of the fluid. Therefore, ${\rm Re}\,e_{1,2}>0$. They lay in the fluid. Condition that   the curve must have no branch points in the fluid (following from incompressibility of the fluid) requires that  two branch points coincide to a real double point: $e_2=e_1=-e_3/2>0$, so that the curve is  degenerate, as in (\ref{16}).  Condition (\ref{g}) yields:
\begin{equation}
e_3=-2\sqrt{-t}\ ,\qquad e_2=e_1=\sqrt{-t}.
\end{equation}
The double point $e_2=e_1$  located in the fluid and the branch point  $e_3$ located  outside  of the fluid are depicted in Figure~\ref{beforeafter}. The curve reads: $Y^2=-(4X^3+12tX+8(-t)^{3/2})$ \cite{us4}.  The uniformizing coordinate is given by the potential $\phi$ by
\begin{equation}
X=e_3-\left(\frac\phi{6}\right)^2\ ,\qquad Y=2\left(\frac{\phi}{6}\right)^3-3e_3\frac{\phi}{6}\ .
\end{equation}
This is equivalent to \eqref{13} if one rescales $\phi$ into $6\phi$. 
The skeleton is a real half axis ${\rm Im}X=0,\;{\rm Re}X<e_3$.

\item
At the critical time $t=0$, all roots coincide $e_3=e_2=e_1=0$. The curve further degenerates to a cusp $Y^2=-4X^3$. The skeleton touches the double point.

\item
After the critical time $t>0$, we push the evolution by splitting the double point $e_2\neq e_1$. The new branch points appear in the fluid, giving rise to shocks.  By the previous argument, we only need to determine the constant term $g_3=4e_3|e_2|^2$.

It follows that the roots and $X$ scale as $t^{1/2}$  and $Y$ scales as $t^{3/4}$. This elliptic curve has no deformation parameters, it is self-similar and in this sense unique.
Scaling properties of the curve and the hydrodynamics are summarized as: 
\begin{align}\label{selfs}
&X(u,t)=|t|^{1/2}X_1(|t|^{1/4}u)\ ,\quad Y(u,t)=|t|^{3/4}Y_1(|t|^{1/4}u)\ , \\
& Y(X,t)=|t|^{3/4}Y_1\left(|t|^{-1/2}X\right),\quad
\phi(X,t)=|t|^{1/4}\phi_1(|t|^{-1/2}X),\\
& \Omega(X,t)=|t|^{5/4}\Omega_1\left(|t|^{-1/2}X\right),\quad
v(X,t)=|t|^{-1/4}v_1\left(|t|^{-1/2}X\right)
\end{align}
where the functions subscripted by $1$ are those evaluated  at $t=\pm 1$.

From the scaling properties alone and from the fact that $Y^2$ is a polynomial of degree three it follows that
\begin{equation}\label{Omega}
\Omega(X)=-{\rm i}\int_{e_3}^X Y\,dX=
-\frac{2\rm i}{5}(XY-2t\phi), 
\end{equation}
\begin{equation}\la{v1}
v(X)=6\ii\frac{X+\frac{C}{12}}{Y(X)}\ ,
\end{equation}
where $\frac{C}{12}=\frac 32\frac{g_3}{g_2}$ is a capacity (see below). We notice that after a critical time velocity diverges at branch points, where $Y=0$.
Eq. (\ref{Omega}) allows to express the total mass deficit accumulated by a shock through the value of  stream function at shock endpoints. Using  $\psi(e_3)=0$  and $Y(e_1)=Y(e_3)=0$, we have:
\be\la{nn}
\frac 1t\int^{e_1}_{e_3}\sigma |\dd X|=\frac 45 |{\rm Re}\,\phi(e_1)|=\frac 45 |\psi(e_1)|.
\ee
The mass scales as $t^{5/4}$. It grows faster than how liquid is getting drained ($\sim t$). We will later show that the rate is a universal constant
\be\la{m}
\frac 1t\int^{e_1}_{e_3}\sigma |\dd X|\approx \frac45 (6.34513)\,t^{1/4}\
\ee

\subsection{Evolution of the elliptic curve}
The Krichever-Boutroux condition (\ref{A2}) uniquely  determines  $g_3$ and therefore  evolution of the curve.  In this case it requires that the integral over $\bf b$-cycle be  purely imaginary,
\begin{equation}\la{KB}
\Omega (e_1)=\mbox{imaginary.}
\end{equation}
It follows from (\ref{Omega})  that this is equivalent with vanishing pressure at branch points:
\be\la{p}
 p(e_{1,2})={\rm Im}\,\phi(e_{1,2})=0.
 \ee
Remarkably, but not accidentally, exactly the same problem appeared in a semiclassical analysis of triply-truncated solution of Painlev\'e I equation \cite{David, Gamba}.

To get the solution, let us parametrize the curve $Y(X)$ by a uniformizing coordinate $u$ as
\begin{align}
X&=\wp(u), \\
Y&={\rm i}\wp'(u),
\end{align}
where $\wp$ is the Weierstrass elliptic function whose half-periods $\omega(t)$ and  $\omega'(t)$ are (yet to be determined) complex functions of time.  Since $e_3$ is real, $\omega+\omega'$ is real,  and $e_2=\bar e_1$, so $\omega=\bar\omega'$.  They scale with time as  $\omega\sim t^{-1/4}$.
The branch points  are given by $(e_3,e_2,e_1)=\left(\wp(\omega+\omega'),\wp(\omega),\wp(\omega')\right)$. The rhombus-shaped  fundamental domain is depicted in  Figure~\ref{rhombuspressure} \cite{abramowitz+stegun}.  As $t$ goes to zero, the rhombus becomes infinitely large, while preserving its aspect ratio.

\begin{center}
\begin{figure}
	\includegraphics[width=8cm]{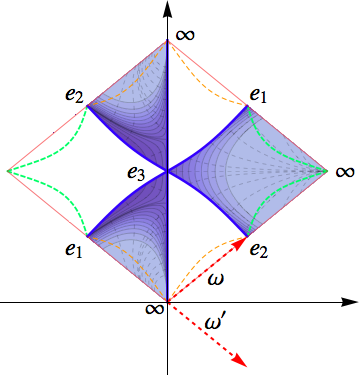} 
	\caption{\label{rhombuspressure} Fundamental domain ($u$-plane).  The pressure level-lines (contour lines) and the stream lines (dashed) are drawn. Pre-image of shaded  regions is a physical plane. 	 Anti-Stokes lines (zero-level lines of ${\rm Re}\,\Omega$) emanate from the branch points.  Among them 
 dashed (orange and green) lines are non-admissible anti-Stokes lines;  the sign of ${\rm Re}\,\Omega$ is different  on different sides of these lines.  The thick (blue) lines are the admissible anti-Stokes lines - shocks.  ${\rm Re}\,\Omega$ is  positive on both sides of these lines.}
\end{figure}
\end{center}

The potential  of the flow is obtained from the Darcy's law by
\begin{equation}\label{phi}
\phi(X)=-\int_{e_3}^X \dd X\,\partial_t Y=6{\rm i}\left(\zeta(u)-\frac{3}{2}\frac{g_3}{g_2}u\right)\ , \qquad g_2=-12 t\ .
\end{equation}
Stream function, $\psi={\rm Re}\,\phi$, and pressure, $p={\rm Im}\,\phi$,  will be read from $\phi(X)$.
Here we have used the facts:
\begin{equation}
u=\ii\int^X\frac{\dd X}{Y(X)}\ ,\qquad \zeta(u)=-\ii\int^X\frac{X \dd X}{Y(X)}\ ,
\end{equation}
up to constants of integration. 

 Eqs. (\ref{p}, \ref{phi}) give the defining equation:
 \begin{equation}\label{phib}
 \frac{3}{2}\frac{g_3}{g_2}=\frac{\zeta(\omega+\omega')}{\omega+\omega'}\ ,\quad \omega'=\bar \omega,
\end{equation} 
whose solution is   summarized as follows:
\begin{align}
\nn&(e_1,e_2,e_3)=\sqrt\frac{12}{4h^2-3}({\textstyle\frac{1}{2}+{\rm i}h,\frac{1}{2}-{\rm i}h},-1)\sqrt t\approx (0.276797+1.79718 \ii,0.276797-1.79718 \ii, -0.553594 )\sqrt t, \\
\nn&g_2=-12\,t, \\
\label{e1e2e3}&g_3=-12\sqrt\frac{12}{4h^2-3}\frac{4h^2+1}{4h^2-3}\,t^{3/2}\approx-7.321762431\,t^{3/2},
\end{align} 
where $h\approx3.246382253744278875676$. This number comes from $m=\frac{1}{2}+\frac{3}{2}\frac{1}{\sqrt{9+4h^2}}$ where $m$ is the solution of the equation 
\begin{equation}\label{EK}
\frac{16m^2-16m+1}{8m^2-9m+1} =\frac{K(m)}{E(m)} ,
\end{equation}
that follows from \eqref{phib}. Here $E$ and $K$ are complete elliptic integrals. They also determine  the {\it invariant} shape of the fundamental domain:
$$\frac{\mbox{Im}\,\omega}{\mbox{Re}\,\omega}=\frac{K'(m)}{K(m)}\approx 0.81736372\ .$$
\end{enumerate}

Eq. \eqref{EK}  follows from \eqref{phib} with a help of  formulas on p. 649 of  \cite{abramowitz+stegun}:
\begin{align}
&\zeta(\omega+\omega')=\frac{K(m)}{3(\omega+\omega')}\big[6E(m)+(4m-5)K(m)\big]\ ,
\\
& g_2=\frac{4(16m^2-16m+1)K^4(m)}{3(\omega+\omega')^4}\ ,
\quad g_3=\frac{8(2m-1)(32m^2-32m-1)K^6(m)}{27(\omega+\omega')^6}\ .
\end{align}

\subsection{Shocks and anti-Stokes lines}
\begin{figure}[htbp]
\begin{center}
\includegraphics[width=6cm]{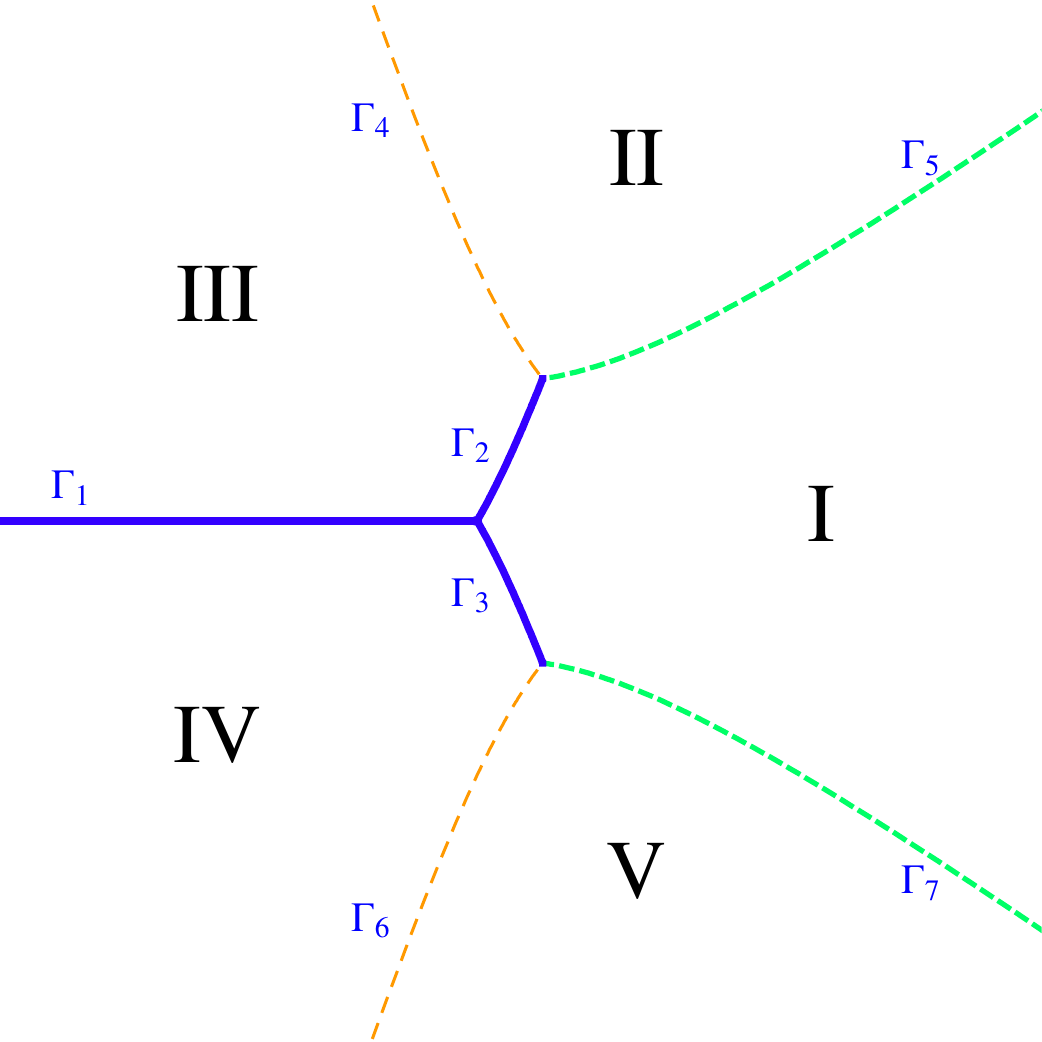}
\caption{Geometry of anti-Stokes lines. The boundary condition  ${\rm Re}\,\Omega >0$ on the remote part of the finger ${\rm arg}\,X=\pi,\;|X|\to \infty$  uniquely determines configuration of shocks $\Gamma_3, \Gamma_2$.}
\label{physicalplane}
\end{center}
\end{figure}

Shock lines are level lines given by  conditions (\ref{AA}, \ref{A2}): 
\begin{equation}
{\rm Re}\,\Omega(X)=0,\quad X\in \Gamma,\quad {\rm Re}\,\Omega(X)>0,\quad X\to \Gamma .
\end{equation}
Condition \eqref{A2} 
\be\la{7}
{\rm Im}\,\big(XY\big)=2t \,{\rm Im}\,\phi(X)=2tp(X),\\
\ee
determines a total of seven anti-Stokes lines connected at the branch points. They are transcendental, computed numerically and denoted by $\Gamma_1,..., \Gamma_7$ in Figure~\ref{physicalplane}.   
Condition (\ref{7}) is valid on both sides of a shock. Since $Y$ changes sign through a shock, so does pressure.

Among the seven anti-Stokes lines, only two $\Gamma_3, \Gamma_2$ are shocks, while $\Gamma_1$ is a skeleton. They are  selected by condition (\ref{AA}). Here is how it works. A  remote part of the finger  ${\rm arg \,X}=\pi,\;|X|\to\infty$ is not affected by the critical transition. It  is a skeleton. Therefore  the branch of $\Omega\sim\frac45X^{5/2}$ must be chosen such that 
${\rm Re}\,\Omega>0$ in both sectors $\{{\rm IV, III}\}$, where at large $X$, $|{\rm arg}\,X-\pi|<2\pi/5$.  This insures that a  line density on the skeleton is positive. It follows that sign of ${\rm Re}\,\Omega$ is 
\begin{equation}
+ ~\mbox{on}~ \{{\rm I, III, IV}\}\ ,\quad - ~\mbox{on}~ \{{\rm II, V}\}\ .
\end{equation}
As a consequence, the signs are opposite on both sides of $\Gamma_4,\Gamma_5,\Gamma_6$ and $\Gamma_7$, and the same (plus) on both sides of $\Gamma_1,\Gamma_2$ and $\Gamma_3$.  The cuts selected this way 
 satisfy the admissibility condition \eqref{AA}. 

\begin{figure}[ht!]
\begin{center}
\includegraphics*[width=5cm]{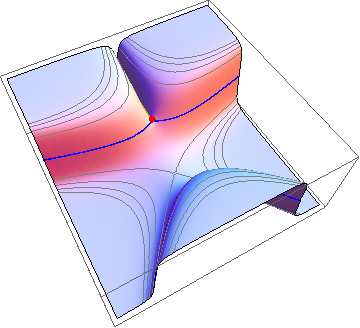}
\includegraphics*[width=5cm]{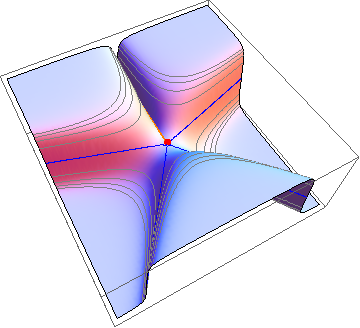}
\includegraphics*[width=5cm]{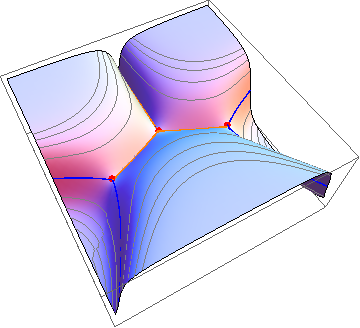}
\caption{Level lines of  ${\rm Re}\,\Omega$  before, at, and after the critical time.  The skeleton and shock lines are marked by (orange) lines (skeleton is hidden under the deep canyon),  the branch points are marked by dots. 
 For visually optimal presentation we  have plotted $\arctan({\rm Re}\,\Omega)$.    }
\label{levels}
\end{center}
\end{figure}

Figure~\ref{levels} illustrates the flow. There,  level lines of ${\rm Re}\, \Omega$ are drawn before, at, and after  the critical time.

The Rankine-Hugoniot conditions (\ref{rh1} and (\ref{7}) give the velocity of shocks. Noting that $\mbox{Im}\,XY=\sigma X_\parallel$, where $X_\parallel$ is a projection of a vector-coordinate of a point of a shock to the shock, we get
\be
V_\perp=\frac{X_\parallel}{t}.
\ee

\subsection{Capacity: discontinuous change of power}\label{capacity1}
 The genus transition is signaled by an abrupt change of $g_3$. It follows from
 \eqref{e1e2e3} that 
\be
g_3=|t|^{3/2}
\left\{
\begin{array}{lr}
-8,&\quad t<0,\\
-7.321762431,&\quad t>0.
\end{array}
\right.
\ee
This discontinuity is related to a  capacity $C$ of the finger which is defined by the asymptote of the potential $\phi$ at $X\rightarrow\infty$ (\ref{YY}):
\begin{equation}\label{Cdef}
\phi(X)=-6 \ii X^{1/2}+\ii\frac{C}{2X^{1/2}}+...
\end{equation}
For an elliptic curve, the capacity  ${C}$ is given by 
\begin{equation}\label{Cdisc}
C=-\dot g_3=\frac32|t|^{1/2}
\left\{
\begin{array}{lr}
-8,\quad &t<0,\\
7.321762431,\quad &t>0.
\end{array}
\right.
\end{equation}

\begin{figure}[htbp]
\begin{center}
\includegraphics[width=6cm]{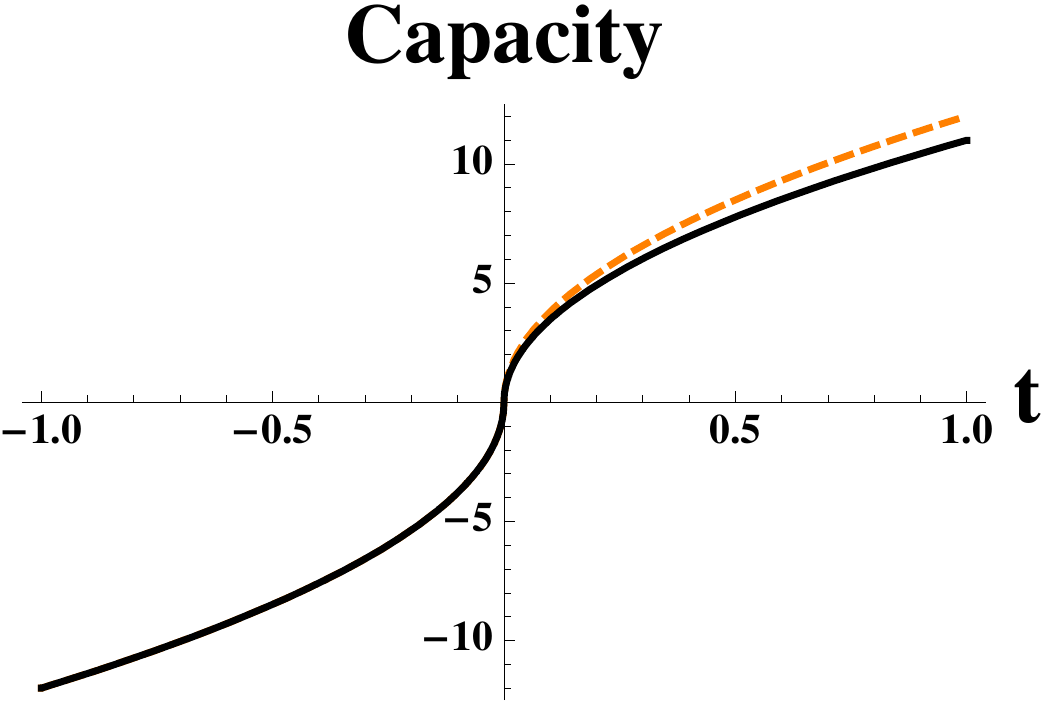}
\caption{Capacity passing the transition (continuous line).  The dashed line corresponds to  no discontinuity at the transition.  The power $N(t)-N_c=-Q^2/(2\pi K)\log {\cal C}$, features the same discontinuity. }
\label{fig_capacity}
\end{center}
\end{figure}

At a branching point, the capacity rate  ${C}/|t|^{1/2}$ goes through a discontinuous jump. Capacity passing through a critical point is drawn on  Figure~\ref{fig_capacity}. Conformal radius of the fluid and  power $N(t)$  also exhibit the discontinuity (see Sec. \ref{N}).

We conjecture that the ratio of the time derivatives of the capacity before and after $1\to 2$ branching, 
\begin{equation}
\eta:=\lim_{t\rightarrow t_c}\frac{\dot { C}_{\mbox{\scriptsize after branching}}}{\dot { C}_{\mbox{\scriptsize before branching}}}
\approx 0.91522030388 \ ,\label{conjecture}
\end{equation}
is universal.
  It does not depend on the details of the flow. A discontinuity  occurs every time when shocks branch \cite{Artem}.
 
 This number can be used to express other objects of the curve.
 Let $\eta:=\frac{3}{2}\frac{g_3}{g_2}t^{-1/2}$ at $t>0$ be:
 $$\eta=\sinh\Theta_c,\quad 
\Theta_c\approx 0.820137.$$
Then, for example, the branch points read:
$$
(e_1,e_2,e_3)\!=\! \left[2\sinh\frac{\Theta_c+\pi\ii}{3},  2\sinh\frac{\Theta_c-\pi\ii}{3}, -2\sinh\frac{\Theta_c}{3}\right]\sqrt t\ .
$$

\subsection{Detailed description of the shock}
\begin{figure}[htbp]
\begin{center}
\includegraphics[width=5cm]{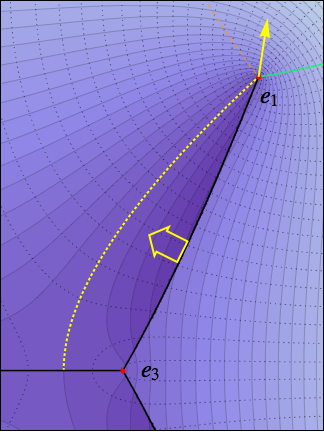}$\quad$
\includegraphics[width=6cm]{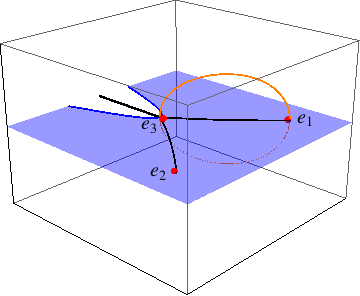}
\caption{The close-up of shocks (left) and the line density profile of the shock $\sigma$  (right). In the upper panel, the shaded contours are equi-pressure lines and the dashed lines are the stream lines.  The darker the shade is, the higher is the pressure in the fluid. The yellow dotted line is the zero-pressure line; it crosses the boundary of the fluid and a skeleton. With respect to this line, the main cut releases/absorbs fluid as $\dot\sigma$ changes the sign. The arrows are moving directions of the shock and of the branch point.   At $e_3$ three directions: zero-pressure line, shock, and the velocity, are all at different angles; see the text. On the left of the zero-pressure line the finger $(x,Y(x))$ expands pushing the fluid away, on the right the finger retreats. 
In the right panel, we plot the line density $\sigma=|Y(X)|$  for one of the shock (orange line in a vertical plane). The line density
 vanishes at  branch points as a square root.  The total charge on the shock grows like $ \frac45 (6.34513)\,t^{5/4}$.  The blue line in the horizontal plane is the line density of  the skeleton, and also a boundary of  the fluid - a viscous finger.}
\label{close}
\end{center}
\end{figure}

It is interesting to follow the zero-pressure line. It always emanates from the shock end points.
 Before the critical time, it is also the boundary of the fluid -- a finger. It is  given by the graph $(x,Y(x))$, where $x={\rm\,Re}\,X$. After the critical time, a zero pressure line has two branches. One remains  the boundary of the fluid, while another  branch is in the fluid and crosses the boundary  at a point where $\dd\phi=0$, as on the upper panel of  Figure~\ref{close}.  This occurs at $x=-\frac{\zeta(\omega+\omega')}{\omega+\omega'}=-\frac32 \frac{g_3}{g_2}=-\eta t^{1/2}$. The zero-pressure line emanates from the branch point $e_1$. If $e_1$ is choosen as origin, the zero-pressure line emanates  with the angle $\theta_{\mbox{\scriptsize zero-pressure}}\approx 0.235\pi$, as we will evaluate below.

The significance of the point $\dd \phi=0$ and the zero-pressure line can be observed in Figure~\ref{close}.  One can see that the stream lines (dotted lines) of fluid that start from the real cut (a skeleton), flow in the opposite directions with respect to the zero-pressure line: the one on the right goes toward the real cut, the one on the left goes away.  Exactly at $\dd \phi=0$, the flow lines are intersecting perpendicularly.
 This means that the boundary of the fluid (a graph $(x,Y(x))$) on the left of this point retreats toward the drain, while the finger expands, as it was before the critical time.  From the right to this point, the boundary moves in the opposite direction, the finger retreats smoothing the tip. The point itself  is a stationary point of fluid where the velocity of  flow and therefore $\nabla_{||}{\rm disc}\,\psi$ vanishes.  It follows from (\ref{weak}) that  $\dot\sigma$ also vanishes at this point.

As has been pointed out, the evolution of the curve $Y(X)$ is self-similar: the whole picture in Figure~\ref{beforeafter} simply expands with respect to the origin by the factor $\sqrt{t}$ for $t>0$ (or $\sqrt{-t}$ for $t<0$).  Therefore, the branching shocks are moving to the darker region (to the left) as time goes on (see the big arrow for the motion of the shock). 

In Figure~\ref{beforeafter}, anti-Stokes lines are located by the condition ${\rm Re}\,\Omega=0$.  The bold lines are admissible anti-Stokes lines selected by the  condition ${\rm Re}\,\Omega>0$ on both sides of the fluid. We have argued that this condition leads to the curl-free flow of the {\it entire fluid} which, after the critical time, consists of decompressed fluid in shocks and a fluid with a constant density  elsewhere.

Let us discuss the  behavior of physical quantities around branch points, say $e_1$. The line density is
\bea
\sigma=|Y| \approx  2|\gamma| \sqrt t \sqrt{|X-e_1|}\ ,
\eea
where $|\gamma|\approx 2.66757$ comes from the estimate below:
\bea\nonumber
\gamma=\frac1t\sqrt{(e_2-e_1)(e_1-e_3)}\approx 2.60534-\ii\, 0.57281.
\eea
From this we can estimate the angle of the shock line $\theta_{\mbox{\scriptsize shock}}$ at its end point $e_1$, by
\begin{equation}\nonumber
\frac32\theta_{\mbox{\scriptsize shock}}+\arg\gamma=\frac{\pi}{2} \quad\Rightarrow \theta_{\mbox{\scriptsize shock}} \approx 0.0459249\pi+\frac{\pi}{3} \ . \end{equation}
Here and below all angles are with respect to the real axis.

The angle between two shocks at their ends is twice $(0.0918498\pi 
+\frac{2\pi}{3})$. It is by $16.532964^o$ larger that the
$\frac{2\pi}{3}$ angle between shocks at their origin.

We also estimate the density near $e_3$ on a skeleton $x<e_3$. It is  $ 
\sigma(x) =|Y|\sim  2\left|\sqrt{-(e_3-e_1)(e_3-e_2)}\right |\sqrt{e_3- 
x}$. It changes in a discontinuous manner  similar to the capacity,  
\eqref{Cdisc}
\begin{align}
\lim_{x\to e_3}&\frac{\sigma_{\mbox{after branching}}} {\sigma_{\mbox{before branching}}} 
=\frac{\left|e_3-e_1\right |}{3\sqrt t} 
\approx 0.659916\ . 
\end{align}
The graph of density is depicted on the lower panel in Figure~\ref{close}.

The total mass deficit carried by a shock follows from (\ref{nn}) and (\ref{phi}) 
 $$\int^{e_1}_{e_3}\sigma |\dd X| =\frac 45 6\left|{\rm Im}\,\left(\zeta(\omega)-\frac{3}{2}\frac{g_3}{g_2}\omega\right)\right |\approx \frac45 (6.34513)\,t^{5/4}\ .$$

Simple calculations give the potential (pressure and stream function) at the branch point
$$
\phi(X)\sim 
e_1\gamma t\sqrt{X-e_1}\approx  (1.7506+\ii 4.5237) \,t \sqrt{X-e_1}\ .
$$

This gives  the angle of the zero-pressure line at $e_1$
\begin{equation}
\frac12\theta_{\mbox{\scriptsize zero-pressure}}+\arg(e_1\gamma)=0\quad\Rightarrow\quad \theta_{\mbox{\scriptsize zero-pressure}}=0.235061\pi\ ,
\end{equation}
such that the angle between the two zero-pressure lines that emanate, respectively, from $e_1$ and $e_2$, is twice the above: $0.470122\pi$, slightly less than $\pi/2$.

Finally, an angle of the  velocity of the branch point $e_1$ (determined by the
 scaling relation (\ref{selfs}) 
$\dot e_k=\frac{1}{2} e_k,\;k=1,2,3$) is  $\theta_{velocity}=0.451357\pi$.

Summing up, velocity of an end point, zero-pressure line and the shock itself at the end point $e_1$ (and symmetrically at $e_2$) all have different directions. The relative angles between them are $\theta_v-\theta_{\mbox{\scriptsize zero pressure}}=0.216296\pi,\;\theta_v-\theta_{\mbox{\scriptsize shock}}=0.0720987\pi$.
 The velocity of the end point is depicted by an arrow in Figure~\ref{close}.

\section{Discussion}

For most initial configurations, zero-surface Hele-Shaw flows evolve into a cusp-singularity in finite time. As such the problem is ill-defined. A singularity signals that the zero-surface Hele-Shaw flow is a singular limit of  a realistic and a well-defined  problem, where velocity and pressure rapidly change at a small scale controlled by some additional small parameters.  In this limit, when these parameters are set to zero, areas of rapidly changing gradients shrink to lines  - shocks. There, velocity and pressure are not differentiable functions and  may even diverge. 

In this paper we found a weak solution of the Hele-Shaw problem, where a few simple physical assumptions almost uniquely determine the evolution of the flow through a singularity, to a growing and branching  shock  graph.

Hele-Shaw flow occurs in a variety of  hydrodynamics settings \cite{Sw, Lipson, Jag-Nag},  aggregation models \cite{DLA81} and  also in   quantum electronic liquids \cite{us8}, random matrices, theory of orthogonal polynomials \cite{us9}  and complex analysis \cite{Wiegmann-Zabrodin00, us2}. In all cases except hydrodynamics a small parameter is available: in aggregation problems it is a size of a aggregating particle, in quantum problems  a small parameter is $\hbar$, in  problems with random matrices and orthogonal polynomial a small parameter is $1/N$, where $N$ is a size of the matrix or an order of a polynomial. In hydrodynamics, ithe varies and depends on a setting. The latter gives a particular (setting-dependent) interpretation of the viscous shock solutions  and requires a case by case study.

In the problem with two fluids (an inviscid fluid pushing a viscous fluid), a shock is a narrow channel filled by a compressed inviscid fluid which forms a line of vortices. In the setting with one viscous fluid (such as a thin layer on a wet surface), sucked away at a distant point, shocks are narrow channels where a fluid layer cracks such as the wet surface (substrate) supplies a turbulent and compressed fluid to the layer.  In both cases, one must relax the condition of incompressibility  and zero-vorticity flow.  Fluids are compressible and not curl-free at the  shock scale. The major physical principle which determines the weak solution is the requirement that the entire fluid (or two fluids) be curl-free and incompressible at a large scale. This means that the smooth part of the flow adjusts to keep the total vorticity zero and a constant draining rate.

In either case, a well-defined set of hydrodynamics equations which may lead to  a shock solution is lacking.

Contrary to a differential Darcy law, the weak solution is not formulated as a Cauchy problem. It relays on an assumption that the boundary fluid is an algebraic curve. In this case, the weak solution uniquely  determines the evolution. Although it seems a rather limited set of initial conditions, the algebraic curve occurs locally at the cusp singularity. In this manner the weak solution describes the evolution through a singularity.  

The condition that the entire flow (a smooth part of the flow and shocks combined) is curl-free brings the problem to an evolution of Krichever-Boutroux curve. These are very special curves which previously appeared in a semiclassical analysis of certain solutions of Painlev\'e I equation  \cite{David, Gamba, Novikov, kapaev}.

In the paper we studied the most generic (2,3) cusp singularity where the Krichever-Boutroux curve is elliptic. In this case the flow produces two shocks.  This solution is universal (parameter-independent) and self-similar. We emphasize interesting universal numbers which describe jumps of physical variables  (like the power $N(t)$  and capacity ${\cal C}(t)$ (\ref{Cdisc})  when the flow goes through a singularity. 

Our solution represents a  local branching event of a  further developed tree.  In Figure~\ref{shocks} we depicted a numerical solution with two generation of branching. 
An unknown, interesting  global structure of  the shock's branching tree is far out of the scope of this paper. It will be very interesting to see whether a developed shock's tree exhibit  a universal scale invariant  limit after a large number of branchings.

\section*{Acknowledgments}

Research of S.-Y L. is supported by CRM-ISM postdoctoral fellowship. P. W was supported by NSF DMR-0540811/FAS 5-27837.
Research or R.~T. was carried out under the auspices of the National Nuclear Security 
Administration of the U.S. Department of Energy at Los Alamos National 
Laboratory under Contract No. DE-AC52-06NA25396. R.~T. acknowledges 
support from the Center for Nonlinear Studies at LANL, and the LDRD Directed Research grant on {\it Physics of Algorithms}.   R.T. and P.W.   acknowledge the hospitality at the Galileo Galilei Institute in Florence, Italy, where  this work was completed.  R.T. also acknowledges the hospitality of the Aspen Center for Physics, and the Centre for Mathematical Research, Montreal, Canada. We thank A. Its, A. Zabrodin, E. Bettelheim and  O. Agam for numerous helpful discussions. P.W. acknowledges discussions with V.  Steinberg, J. Feinberg and E. Sharon on cracks observed in visco-elastic media, and H. Jaeger and S. Nagel  on Hele-Shaw flow in granular media.
We are especially grateful to I. Krichever and  M. Bertola for discussions and sharing their results about Boutroux curves and Ar. Abanov for his important comments and insightful discussions. 
\def\cprime{$'$} \def\cprime{$'$} \def\cprime{$'$} \def\cprime{$'$}
  \def\cprime{$'$}


\begin{thebibliography}{10}

\bibitem{Hele-Shaw}
H.~S.~S. Hele-Shaw.
\newblock {\em Nature}, 58(1489):34--36, 1898.

\bibitem{Darcy}
H.~Darcy.
\newblock {\em Fontaines publiques de la ville de {D}ijon}.
\newblock Librairie des Corps Imp\'eriaux des Ponts et Chauss\'ees et des
  Mines, Paris, 1856.

\bibitem{Hohlov-Howison94}
Y.~E. Hohlov and S.~D. Howison.
\newblock On the classification of solutions to the zero-surface-tension model
  for {H}ele-{S}haw free boundary flows.
\newblock {\em Quart. Appl. Math.}, 51(4):777--789, 1993.

\bibitem{Gustafsson-Vasiliev06}
B.~Gustafsson and A.~Vasil{\cprime}ev.
\newblock {\em Conformal and potential analysis in {H}ele-{S}haw cells}.
\newblock Advances in Mathematical Fluid Mechanics. Birkh\"auser Verlag, Basel,
  2006.

\bibitem{ST}
P.~G. Saffman and G.~Taylor.
\newblock The penetration of a fluid into a porous medium or {H}ele-{S}haw cell
  containing a more viscous liquid.
\newblock {\em Proc. Roy. Soc. London. Ser. A}, 245:312--329. (2 plates), 1958.

\bibitem{bs84}
B.~Shraiman and D.~Bensimon.
\newblock Singularities in nonlocal interface dynamics.
\newblock {\em Phys. Rev. A}, 30(5):2840--2842, 1984.

\bibitem{Howison85}
S.~D. Howison, J.~R. Ockendon, and A.~A. Lacey.
\newblock Singularity development in moving-boundary problems.
\newblock {\em Quart. J. Mech. Appl. Math.}, 38(3):343--360, 1985.

\bibitem{Richardson72}
S.~Richardson.
\newblock {Hele {S}haw flows with a free boundary produced by the injection of
  fluid into a narrow channel}.
\newblock {\em Journal of Fluid Mechanics}, 56:609--618, 1972.

\bibitem{us1}
M.~Mineev-Weinstein, P.~B. Wiegmann, and A.~Zabrodin.
\newblock Integrable structure of interface dynamics.
\newblock {\em Physical Review Letters}, 84:5106, 2000.

\bibitem{us3}
I.~Krichever, M.~Mineev-Weinstein, P.~Wiegmann, and A.~Zabrodin.
\newblock Laplacian growth and {W}hitham equations of soliton theory.
\newblock {\em Physica D}, 198:1, 2004.

\bibitem{us4}
R.~Teodorescu, A.~Zabrodin, and P.~Wiegmann.
\newblock Unstable fingering patterns of {H}ele-{S}haw flows as a
  dispersionless limit of the {K}d{V} hierarchy.
\newblock {\em Physical Review Letters}, 95:044502, 2005.

\bibitem{us5}
E.~Bettelheim, O.~Agam, A.~Zabrodin, and P.~Wiegmann.
\newblock Singular limit of {H}ele-{S}haw flow and dispersive regularization of
  shock waves.
\newblock {\em Physical Review Letters}, 95:244504, 2005.

\bibitem{us6}
S-Y. Lee, E.~Bettelheim, and P.~Wiegmann.
\newblock Bubble break-off in {H}ele-{S}haw flows : Singularities and
  integrable structures.
\newblock {\em Physica D}, 219:22, 2006.

\bibitem{Sw}
E.~Sharon, M.~G. Moore, W.~D. McCormick, and H.~L. Swinney.
\newblock Coarsening of fractal viscous fingering patterns.
\newblock {\em Phys. Rev. Lett.}, 91(20):205504, 2003.

\bibitem{Maher}
H.~Zhao and J.~V. Maher.
\newblock Associating-polymer effects in a {H}ele-{S}haw experiment.
\newblock {\em Phys. Rev. E}, 47(6):4278--4283, 1993.

\bibitem{Lipson}
S.~G. Lipson.
\newblock Pattern formation in drying water films.
\newblock {\em Physica Scripta}, T67:63--66, 1996.

\bibitem{Jag-Nag}
X.~{Cheng}, L.~{Xu}, A.~{Patterson}, H.~M. {Jaeger}, and S.~R. {Nagel}.
\newblock {Towards the zero-surface-tension limit in granular fingering
  instability}.
\newblock {\em Nature Physics}, 4:234--237, March 2008.

\bibitem{DLA81}
T.~A. Witten and L.~M. Sander.
\newblock Diffusion-limited aggregation, a kinetic critical phenomenon.
\newblock {\em Phys. Rev. Lett.}, 47(19):1400--1403, 1981.

\bibitem{HastingsLevitov}
M.~B. Hastings and L.~S. Levitov.
\newblock Laplacian growth as one-dimensional turbulence.
\newblock {\em Physica D 116}, 244, 1998.

\bibitem{LL}
L.~D. Landau and E.~M. Lifshits.
\newblock {\em Fluid {M}echanics}.
\newblock Butterworth-{H}einemann, 1987.

\bibitem{Ahlfors}
L.~V. Ahlfors.
\newblock {\em Complex analysis: {A}n introduction of the theory of analytic
  functions of one complex variable}.
\newblock Second edition. McGraw-Hill Book Co., New York, 1966.

\bibitem{Artem}
Ar. Abanov.
\newblock {\em private communication}.

\bibitem{Evans}
L.C. Evans.
\newblock {\em Partial differential equations}, volume~19 of {\em Graduate
  Studies in Mathematics}.
\newblock American Mathematical Society, Providence, RI, 1998.

\bibitem{Lamb}
H.~Lamb.
\newblock {\em Hydrodynamics}.
\newblock Cambridge University Press, Cambridge, 1993.

\bibitem{PK}
P.~Ya. Polubarinova-Kochina.
\newblock {\em Dokl. Acad. Nauk SSSR}, 47:254--7, 1945.

\bibitem{ms}
M.~B. Mineev-Weinstein and S.~P. Dawson.
\newblock Class of nonsingular exact solutions for {L}aplacian pattern
  formation.
\newblock {\em Phys. Rev. E}, 50(1):R24--R27, 1994.

\bibitem{Bell04}
S.~R. Bell.
\newblock The {B}ergman kernel and quadrature domains in the plane.
\newblock In {\em Quadrature domains and their applications}, volume 156 of
  {\em Oper. Theory Adv. Appl.}, pages 61--78. Birkh\"auser, Basel, 2005.

\bibitem{Galin}
L.~A. Galin.
\newblock {\em Dokl. Acad. Nauk SSSR}, 47(1-2):250--3, 1945.

\bibitem{Howison86}
S.~D. Howison.
\newblock Cusp development in {H}ele-{S}haw flow with a free surface.
\newblock {\em SIAM J. Appl. Math.}, 46(1):20--26, 1986.

\bibitem{Mineev90}
M.~B. Mineev.
\newblock A finite polynomial solution of the two-dimensional interface
  dynamics.
\newblock {\em Physica D}, 43(2-3):288--292, 1990.

\bibitem{Shraiman}
B.~I. Shraiman.
\newblock Velocity selection and the {S}affman-{T}aylor problem.
\newblock {\em Phys. Rev. Lett.}, 56(19):2028--2031, May 1986.

\bibitem{King}
J.~R. King, A.~A. Lacey, and J.~L. V{\'a}zquez.
\newblock Persistence of corners in free boundaries in {H}ele-{S}haw flow.
\newblock {\em European J. Appl. Math.}, 6(5):455--490, 1995.
\newblock Complex analysis and free boundary problems (St.\ Petersburg, 1994).

\bibitem{us9}
R.~Teodorescu, E.~Bettelheim, O.~Agam, A.~Zabrodin, and P.~Wiegmann.
\newblock Normal random matrix ensemble as a growth problem.
\newblock {\em Nuclear Physics B}, 704:407, 2005.

\bibitem{S3}
M.~Sakai.
\newblock Regularity of a boundary having a {S}chwarz function.
\newblock {\em Acta Math.}, 166(3-4):263--297, 1991.

\bibitem{Davis}
P.~J. Davis.
\newblock {\em The {S}chwarz function and its applications}.
\newblock The Mathematical Association of America, Buffalo, N. Y., 1974.
\newblock The Carus Mathematical Monographs, No. 17.

\bibitem{G4}
B.~Gustafsson.
\newblock Lectures on balayage.
\newblock In {\em Clifford algebras and potential theory}, volume~7 of {\em
  Univ. Joensuu Dept. Math. Rep. Ser.}, pages 17--63. Univ. Joensuu, Joensuu,
  2004.

\bibitem{Krichev-red}
I.~M. Krichever.
\newblock The {$\tau$}-function of the universal {W}hitham hierarchy, matrix
  models and topological field theories.
\newblock {\em Comm. Pure Appl. Math.}, 47(4):437--475, 1994.

\bibitem{David} F. ~David, \newblock  Phases of the large-N matrix model and non-perturbative effects in 2D gravity
\newblock {\em Nuclear Physics B}, 348(3): 507-524, 1991;
\newblock Non-perturbative effects in matrix models and vacua of two dimensional gravity, \newblock {\em Physics Letters B}
302(4): 403-410, 1993;  

\bibitem{Bertola-Mo}
M.~Bertola and M.~Y. Mo.
\newblock Commuting difference operators, spinor bundles and the asymptotics of
  pseudo-orthogonal polynomials with respect to varying complex weights.
\newblock {\em [arXiv.org:math-ph/0605043]}, 2006.

\bibitem{Gamba}
F.~Fucito, A.~Gamba, M.~Martellini, and O.~Ragnisco.
\newblock Non-linear {WKB} analysis of the string equation.
\newblock {\em International Journal of Modern Physics B}, 6:2123, 1992.

\bibitem{Novikov}
P.~G. Grinevich and S.~P. Novikov.
\newblock String equation--2. physical solution.
\newblock {\em St. Petersburg Math. J.}, 6:553, 1995.

\bibitem{abramowitz+stegun}
M.~Abramowitz and I.~A. Stegun.
\newblock {\em Handbook of Mathematical Functions with Formulas, Graphs, and
  Mathematical Tables}.
\newblock Dover, New York, 1964.

\bibitem{us8}
O.~Agam, E.~Bettelheim, P.~Wiegmann, and A.~Zabrodin.
\newblock Viscous fingering and a shape of an electronic droplet in the
  {Q}uantum {H}all regime.
\newblock {\em Physical Review Letters}, 88:236801, 2002.

\bibitem{Wiegmann-Zabrodin00}
P.~B. Wiegmann and A.~Zabrodin.
\newblock Conformal maps and integrable hierarchies.
\newblock {\em Comm. Math. Phys.}, 213(3):523--538, 2000.

\bibitem{us2}
I.~K. Kostov, I.~Krichever, M.~Mineev-Weinstein, P.~Wiegmann, and A.~Zabrodin.
\newblock {$\tau$}-function for analytic curves [arxiv.org:hep-th/0005259].
\newblock {\em MSRI Publications}, 40:285, 2001.

\bibitem{kapaev}
A.~A. Kapaev.
\newblock Monodromy deformation approach to the scaling limit of the {P}ainleve
  first equation.
\newblock {\em CRM Proc. Lecture Notes}, 32:157, 2002.

\end{thebibliography}
\end{document}